%% file: rsz_temperature.tex
\documentclass[usenatbib,a4paper]{mn2e}
\usepackage{amsmath}
\usepackage{amssymb}
\usepackage{amsfonts}
\usepackage {multirow}
\usepackage {graphicx}
\usepackage {natbib}
\usepackage{txfonts}
\usepackage{graphicx}
\usepackage[hyperindex,breaklinks=true, colorlinks, citecolor=blue]{hyperref}
\usepackage{tablefootnote}
\usepackage{eurosym}
\usepackage{ulem}
\usepackage[usenames,dvipsnames]{xcolor}

\input{Planck.tex}

\newcommand{\tC}{\boldsymbol{\rm C}}

\newcommand{\Pe}{P_{\rm e}}
\newcommand{\Ne}{n_{\rm e}}
\newcommand{\Te}{T_{\rm e}}
\newcommand{\Tewh}{\widehat{T}_{\rm e}}
\newcommand{\Teol}{\overline{T}_{\rm e}}
\newcommand{\me}{m_{\rm e}}
\newcommand{\vek}[1]{\boldsymbol{#1}}

\def\reff@jnl#1{{\rm#1\/}}

\def\aj{\reff@jnl{AJ}}                  
\def\araa{\reff@jnl{ARA\&A}}            
\def\apj{\reff@jnl{ApJ}}                
\def\apjl{\reff@jnl{ApJ}}               
\def\apjs{\reff@jnl{ApJS}}              
\def\ao{\reff@jnl{Appl.Optics}}         
\def\apss{\reff@jnl{Ap\&SS}}            
\def\aap{\reff@jnl{A\&A}}               
\def\aapr{\reff@jnl{A\&A~Rev.}}         
\def\aaps{\reff@jnl{A\&AS}}             
\def\azh{\reff@jnl{AZh}}                        
\def\baas{\reff@jnl{BAAS}}              
\def\jcap{\reff@jnl{JCAP}}              
\def\jrasc{\reff@jnl{JRASC}}            
\def\memras{\reff@jnl{MmRAS}}           
\def\mnras{\reff@jnl{MNRAS}}            
\def\pra{\reff@jnl{Phys.Rev.A}}         
\def\prb{\reff@jnl{Phys.Rev.B}}         
\def\prc{\reff@jnl{Phys.Rev.C}}         
\def\prd{\reff@jnl{Phys.Rev.D}}         
\def\prl{\reff@jnl{Phys.Rev.Lett}}      
\def\pasp{\reff@jnl{PASP}}              
\def\pasj{\reff@jnl{PASJ}}              
\def\qjras{\reff@jnl{QJRAS}}            
\def\skytel{\reff@jnl{S\&T}}            
\def\solphys{\reff@jnl{Solar~Phys.}}    
\def\sovast{\reff@jnl{Soviet~Ast.}}     
 \def\ssr{\reff@jnl{Space~Sci.Rev.}}    
\def\zap{\reff@jnl{ZAp}}                
\def\nat{\reff@jnl{Nature}}             
\def\procspie{\reff@jnl{Proceedings of the SPIE}}             

\def\Planck{\textit{Planck}}

\def\litebird{\textit{LiteBIRD}}

\def\pico{\textit{PICO}}

\def\ben{\begin{enumerate}}
\def\een{\end{enumerate}}
\def\bi{\begin{itemize}}
\def\ei{\end{itemize}}
\def\be{\begin{equation}}
\def\ee{\end{equation}}
\def\bea{\begin{eqnarray}}
\def\eea{\end{eqnarray}}
\def\ba{\begin{align}}
\def\ea{\end{align}}

\def\bdw{\boldsymbol{w }}
\def\bda{\boldsymbol{a }}

\def\bde{\boldsymbol{e }}

\def\bdn{\boldsymbol{n }}

\def\bdx{\boldsymbol{x }}
\def\bdw{\boldsymbol{w }}

\def\bdf{\boldsymbol{f }}

\newcommand{\tA}{\boldsymbol{{\rm A}}}

\newcommand{\healpix}{{\tt HEALPix}}

\makeatletter
\newcommand\footnoteref[1]{\protected@xdef\@thefnmark{\ref{#1}}\@footnotemark}
\makeatother

\title[Mapping SZ cluster temperatures]
{Mapping the relativistic electron gas temperature across the sky}
\author[Mathieu Remazeilles and Jens Chluba]{Mathieu Remazeilles\thanks{E-mail:~\href{mailto:mathieu.remazeilles@manchester.ac.uk}{\textcolor{black}{mathieu.remazeilles@manchester.ac.uk}}} and Jens Chluba
\\
Jodrell Bank Centre for Astrophysics, School of Physics and Astronomy, The University of Manchester, Oxford Road, Manchester, M13 9PL, U.K.
}

\begin{document}

\date{{Accepted  --. Received }}

\maketitle

\begin{abstract}
With increasing sensitivity, angular resolution, and frequency coverage, future cosmic microwave background (CMB) experiments like \pico\ will allow us to access new information about galaxy clusters through the \textit{relativistic} thermal Sunyaev-Zeldovich (SZ) effect. We will be able to map the temperature of relativistic electrons across the entire sky, going well beyond a simple detection of the relativistic SZ effect by cluster stacking methods that currently define the state-of-the-art. Here, we propose a new map-based approach utilizing SZ-temperature \textit{moment expansion} and \textit{constrained-ILC} methods to extract electron gas temperature maps from foreground-obscured CMB data. This delivers a new independent map-based observable, the electron temperature power spectrum  $\Te^{yy}(\ell)$, 
which can be used to constrain cosmology in addition to the Compton-$y$ power spectrum $C_\ell^{yy}(\ell)$ . We find that \pico\ has the required sensitivity, resolution, and frequency coverage to accurately map the electron gas temperature of galaxy clusters across the full sky, covering a broad range of angular scales. 
Frequency-coverage at $\nu\gtrsim 300\,{\rm GHz}$ plays an important role for extracting the relativistic SZ effect in the presence of foregrounds. For Coma, \pico\ will allow us to directly reconstruct the electron temperature profile using the relativistic SZ effect. Coma's average electron temperature will be measured to $10\sigma$ significance after foreground removal using \pico. Low-angular resolution CMB experiment like \litebird\ could achieve $2\sigma$ to $3\sigma$ measurement of the electron temperature of this largest cluster. Our analysis highlights a new spectroscopic window into the thermodynamic properties of galaxy clusters and the diffuse electron gas at large angular scales.
\end{abstract}


\section{Introduction}
\label{sec:intro}

The energy spectrum of the cosmic microwave background (CMB) radiation is long known to be extremely close to a perfect blackbody since the measurements by COBE/FIRAS \citep{Mather1994, Fixen1996}. Nevertheless, several processes release energy to the CMB photons throughout the thermal history of the Universe, thereby imprinting tiny frequency-dependent signals that are commonly referred to as CMB spectral distortions \citep[e.g.,][]{Chluba2011therm, Sunyaev2013, Tashiro2014, deZotti2015, Chluba2016, Chluba2019Dec}. One of the most significant mechanism is the so-called thermal Sunyaev-Zeldovich (tSZ) effect \citep{Zeldovich1969, Sunyaev1972}, which imprints an anisotropic spectral distortion from the low-redshift Universe due to inverse Compton scattering of CMB photons by hot electrons residing inside galaxy clusters \citep[e.g.,][]{Birkinshaw1999, Carlstrom2002, SZreview2019}. The resulting tSZ $y$-type distortion (i.e., Comptonization) of the CMB spectrum at the positions of galaxy clusters in the sky has become a powerful tool for detecting new galaxy clusters \citep{Planck2015_XXVII}, measuring their thermodynamic properties \citep{Planck_int_resultsV2013}, and constraining cosmological parameters such as the amplitude of dark matter fluctuations \citep[e.g.,][]{Planck2015ymap,Planck2015clustercount}.

Mapping the thermal SZ Compton-$y$ parameter over the sky has become a routine exercise with third-generation CMB experiments \citep{Planck2015ymap,PACT2019}, offering a direct probe the electron gas pressure $\Pe = \Ne k \Te$ of galaxy clusters integrated along the line-of-sight:
\begin{align}
\label{eq:y}
y = \frac{\sigma_T}{\me c^2} \int \Ne(l)\, k\Te(l)\,{\rm d}l,
\end{align}
where $\sigma_T$ is the Thomson scattering cross-section, $\me$ the electron rest mass, $c$ the speed of light, $k$ the Boltzmann constant, $l$ the line-of-sight coordinate, $\Ne$ the electron density in the galaxy cluster, and $\Te$ the temperature of the electron gas. To date, SZ measurements have not allowed to disentangle electron density and temperature because of their degeneracy in the Compton-$y$ parameter Eq.~\eqref{eq:y}, so that we still have to rely on X-ray measurements of galaxy clusters to probe the cluster temperatures $\Te^X \simeq \Te$ \citep{Arnaud2005,Pratt2007}. However, X-rays trace the central denser component of the cluster by scaling as $\Ne^2$ unlike the thermal SZ effect which scales as $\Ne$, but the cluster gas is not homogeneous nor isothermal, therefore X-ray temperatures are not the best proxy for the actual virial temperatures of galaxy clusters, in contrast to SZ temperatures \citep{Pointecouteau1998,Kay2008,Lee2019}.

Future ground-based CMB experiments like the Simons Observatory \citep[SO;][]{SO2019} will offer high spatial resolution across the frequency range $27$-$280$\,GHz to detect more than $10^4$ galaxy clusters through SZ effect and explore broad cluster science. However, as illustrated in Fig.~\ref{Fig:seds}, high frequencies above $300$\,GHz greatly help in breaking the temperature degeneracies in the relativistic thermal SZ spectrum present at lower frequency \citep[e.g.,][]{Chluba2012moments,Mittal2018, Erler2018, Astro20202019Basu,voyage2050_backlight}.
As we illustrate here, with increasing sensitivity and frequency coverage, next-generation CMB space missions like \pico\ \citep{PICO2019}  and \litebird\ \citep{Suzuki2018} will enable us to measure the faint \textit{relativistic} corrections to the thermal SZ effect, whose spectral signature depends on the electron temperature \citep{Wright1979, Rephaeli1995, Sazonov1998, Challinor1998,Itoh1998}. This should allow us in principle to extract the electron temperature and density profiles of galaxy clusters in the outskirts, beyond the radius of virialization, and thus to probe the missing baryons in the Universe without relying on external X-ray measurements. While the first "SZ revolution" has seen the mapping of Compton-$y$ parameter and electron pressure of galaxy clusters across the sky; a second "SZ revolution" will emerge from next-generation CMB experiments with the mapping of the electron temperatures and densities across the sky, advancing our knowledge of the thermodynamic properties of galaxy clusters and offering new map-based observables for cosmology. In this work, we develop a new map-based methodology to achieve this goal with future CMB observations.

The intensity of the thermal SZ effect at frequency $\nu$ and at the position $\vek{\theta}=(\theta_1,\theta_2)$ in the spherical sky depends on the local electron gas temperature $\Te$ as:
\begin{align}
\label{eq:tSZ}
\Delta I^{\rm tSZ}\big(\nu,\vek{\theta},\Te(\vek{\theta})\big) = f\big(\nu,\Te(\vek{\theta})\big)\, y(\vek{\theta}),
\end{align}
where the Compton parameter $y(\vek{\theta})$ accounts for the spatial fluctuations of the thermal SZ intensity across the sky, while the energy spectrum $f(\nu,\Te)$ account for the spectral variations of the thermal SZ intensity both across frequencies and across the sky, since the cluster temperatures $\Te=\Te(\vek{\theta})$ vary across the sky. In the \textit{non-relativistic} limit, which is valid only for very low temperatures, $k\Te/\me c^2 \ll 1$, the SZ spectrum reads \citep{Zeldovich1969,Sunyaev1972}
\begin{align}
\label{eq:non-rel}
f\big(\nu,\Te\simeq 0\big) = \frac{2h}{c^2} \left(\frac{kT_{\rm CMB}}{h}\right)^3 \frac{x^4 {\rm e}^x}{({\rm e}^x-1)^2} \left[x\,{\rm coth}\left(\frac{x}{2}\right)\,-\,4\right], 
\end{align}
where $c$ is the speed of light, $\me$ is the electron rest mass, and ${x\equiv h\nu / kT_{\rm CMB}}$ with $h$ being the Planck constant, $k$ the Boltzmann constant, and $T_{\rm CMB}$ the CMB blackbody temperature. 

However, galaxy clusters are massive (with typical masses $M \gtrsim 3\times 10^{14}\,h^{-1}M_{\sun}$) and thus hot, with a typical average temperatures of $k\Teol \gtrsim 5$\,keV \citep{Refregier2000,Komatsu:2002wc,Erler2018,Remazeilles2019}. This means that the electrons in the virialized gas hosted by galaxy clusters are relativistic, with thermal velocities approaching a significant fraction of the speed of light $\varv_{\rm th} = \sqrt{2k\Te/\me}\simeq 0.1$-$0.2c$. Hence, the full \textit{relativistic} description of the thermal SZ effect (Eq.~\ref{eq:tSZ}) has to be adopted for accurate astrophysical and cosmological analyses of galaxy clusters \citep{Hurier2017rSZ, Erler2018,Remazeilles2019}.

In this work, we aim at mapping the cluster temperatures $\Te(\vek{\theta})$ across the sky, by disentangling relativistic SZ from non-relativistic SZ effects in foreground-contaminated CMB data. Our approach is to perform an SZ-temperature moment expansion of the thermal SZ intensity \citep{Chluba2012moments} to discriminate different spectral components, $y(\vek{\theta})$ and $\Te(\vek{\theta})$, contributing to the thermal SZ emission. We then use the Constrained-ILC method \citep{Remazeilles2011} for component separation in order to fully disentangle the spatially correlated fields $\Te(\vek{\theta})$ and $y(\vek{\theta})$. 

Previously, \cite{Hurier2017rSZ} have proposed a similar approach to the Constrained-ILC for extracting relativistic SZ effects with \textit{COrE+}. 
However, they used a secant approximation for the temperature derivative of the SZ spectrum, instead of the exact derivative as we do here. Pioneering works aiming at modelling the relativistic SZ corrections are based on asymptotic expansions \citep[e.g.][]{Itoh1998, Challinor1998}, and as such are equivalent to using a \textit{zero} pivot temperature in Eq.~\eqref{eq:moment} for any cluster. This is a poor approximation for most clusters, and biases the measured $y$ and $\Te$ fields (see Fig.~\ref{Fig:all-sky-te} and Fig.~\ref{Fig:average_te}).
The novelty of our method is to determine in first instance the appropriate pivot temperature for any cluster by performing a \textit{temperature spectroscopy} of the cluster (see Fig.~\ref{Fig:modulated} and Fig.~\ref{Fig:all-sky}), then to perform component separation around the relevant pivot temperature in the next iteration. In addition, we deproject kinetic SZ (kSZ) effect (and as a byproduct CMB temperature fluctuations), since the kSZ (velocity) profile of an individual cluster can bias the relativistic temperature profile of this cluster, and as such is a significant foreground to the relativistic SZ effect. 
We also add an extra constraint on the zeroth moment of thermal dust emission to mitigate the bulk of this foreground contamination in the reconstructed temperature field. Finally, envisioning next-generation CMB experiments like \pico\ and \litebird, we perform for the first time a reconstruction of the electron gas temperature across the entire sky, delivering full-sky $\Te$-maps along with the $y$-maps and deriving the power spectrum of the $y^2$-weighted electron temperature, hence advocating a novel map-based observable, $\Te^{yy}(\ell)$, in addition to the usual $y$-map power spectrum to constrain cosmology with future SZ observations.

This paper is organised as follows. The formalism of our component separation method to extract SZ temperatures is described in Sect.~\ref{sec:temp}. Our sky simulations of the \pico\ experiment are explained in Sect.~\ref{sec:simu}. We perform the analysis of the sky simulations and present our results in Sect.~\ref{sec:analysis}. We conclude in Sect.~\ref{sec:conc}.

\vspace{-3mm}
\section{Mapping cluster temperatures}
\label{sec:temp}

\subsection{Component separation}
\label{subsec:method}

The moment expansion of the relativistic thermal SZ intensity (Eq.~\ref{eq:tSZ}) around some pivot temperature $\Teol$ is \citep{Chluba2012moments}
\begin{align}
\label{eq:moment}
\Delta I^{\rm tSZ}\big(\nu,\vek{\theta},\Te\big) 
&\!\simeq \!f\left(\nu, \Teol\right) y\left(\vek{\theta}\right) + \frac{\partial f\left(\nu,\Teol\right)}{\partial \Teol} y\left(\vek{\theta}\right)\, \left(\Te\left(\vek{\theta}\right) - \Teol\right),
\end{align}
where higher-order terms are neglected, assuming that the pivot temperature $\Teol$ is chosen to be as close as possible to the \textit{average} temperature of galaxy cluster\footnote{Even within a cluster or for ensembles of many clusters across the sky, temperature dispersion contributes to the exact SZ signal \citep{Chluba2012moments}. For higher precision this mean adding higher order moment terms. However, here we neglect these corrections.}, $\Teol^y = \langle y\,\Te \rangle\,/\, \langle y \rangle$.

\begin{figure}
  \begin{center}
    \includegraphics[width=\columnwidth]{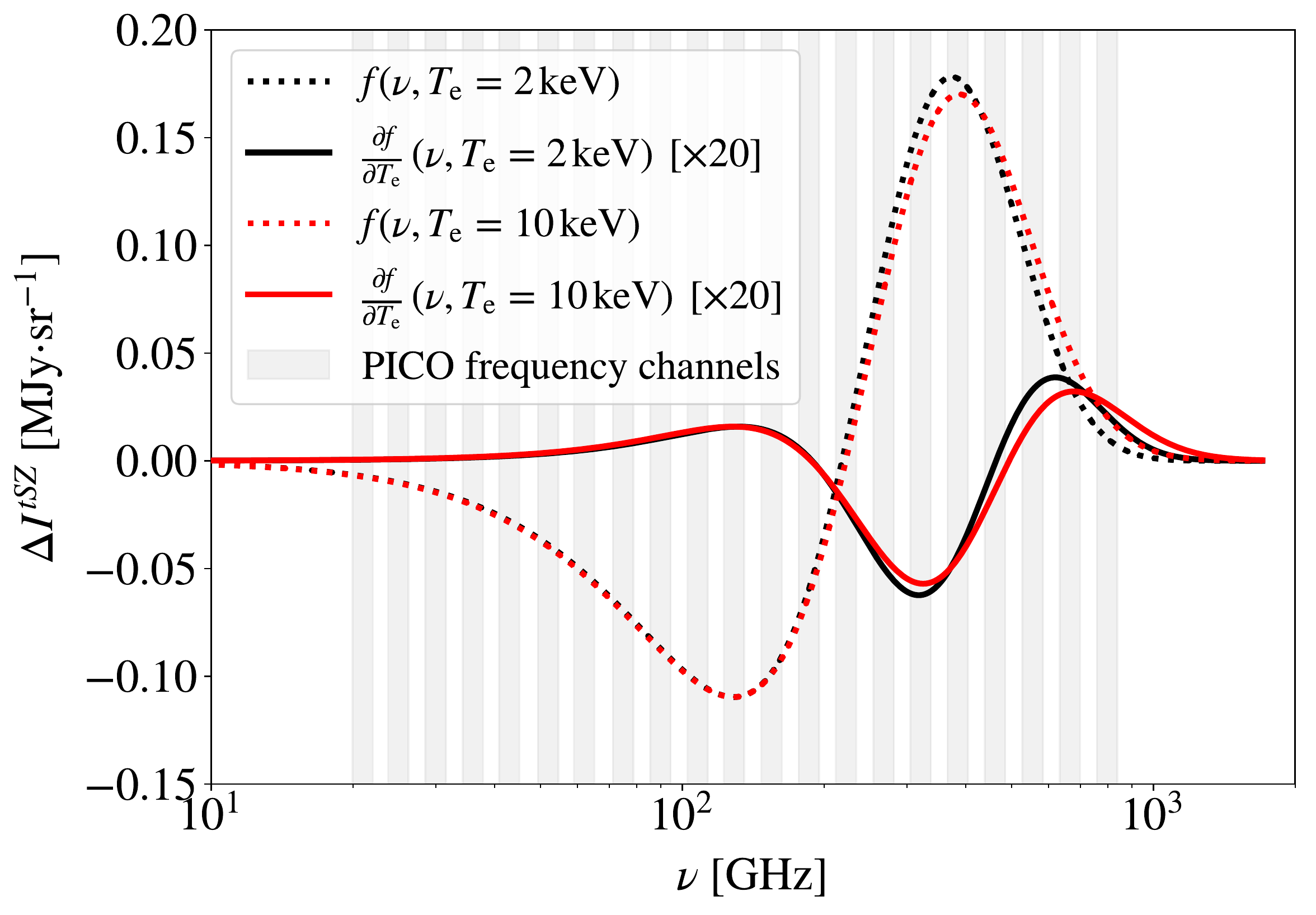}
      \end{center}
\caption{{Spectral energy distribution of relativistic thermal SZ components for different temperatures: $\Teol=2$\,keV  (\textit{black}) and $\Teol=10$ \,keV (\textit{red}).} \textit{Dotted lines}: spectrum $f(\nu,\Teol)$ of the relativistic Compton-$y$ component. \textit{Solid lines}: spectrum ${\partial f(\nu,\Teol) / \partial \Teol}$ (enhanced by a factor $20$) of the temperature-modulated ${y(\Te - \Teol)}$ component. A varying Compton-$y$ is used to rescale 
the amplitude of the spectra so that they match  at low frequencies. This highlights the 
spectral shape degeneracy (i.e. $y$-$\Te$ degeneracy) in the thermal SZ spectrum at low frequency for any temperature, and the importance of high frequencies $\gtrsim 300$\,GHz to break this degeneracy  and distinguish between different temperatures. Grey areas are \pico\ frequency bands with reduced $5$\% bandwidth for clarity.}
\label{Fig:seds}
\end{figure}

Sky observations from CMB experiments are obscured by Galactic foregrounds and instrumental noise, so that the total intensity of the sky emission can be modelled as:
\begin{align}
\label{eq:model}
\Delta I\left(\nu,\vek{\theta}\right)  \simeq f\left(\nu, \Teol\right)\, y(\theta)\, +\, \frac{\partial f\left(\nu,\Teol\right)}{ \partial \Teol}\,y(\theta)\, \left(\Te(\theta) - \Teol\right)\,+\, n\left(\nu,\vek{\theta}\right),
\end{align}
where $n\left(\nu,\vek{\theta}\right)$ accounts for the total \textit{nuisance} emission from foregrounds and noise, i.e. everything that is not due to thermal SZ effects. It is worth noting that at this point we do not attempt to parametrise the spectral properties of the nuisance term, keeping this model as blind as possible concerning poorly known Galactic foregrounds. We will augment the model with partial constraints on dust and kSZ below (see Eq.~\ref{eq:constbis}). In Eq.~\eqref{eq:model}, the Compton parameter signal, $y(\vek{\theta})$, and the \textit{temperature-modulated} Compton parameter signal, ${y(\vek{\theta})\, (\Te(\vek{\theta}) - \Teol)}$, can be regarded as two different components of emission with two distinct spectral signatures, respectively $f(\nu, \Teol)$ and ${\partial f(\nu,\Teol) / \partial \Teol}$ (Fig.~\ref{Fig:seds}). Therefore, it is in principle possible to disentangle the two SZ signals through multi-frequency observations and component separation methods.

While the two components of the SZ signal, $y$ and ${y(\Te - \Teol)}$, are spectrally decorrelated, they are strongly spatially correlated. Therefore, we will use the \textit{Constrained-ILC} component separation method \citep{Remazeilles2011} to project the multi-frequency data orthogonally to the energy spectrum, $f(\nu, \Teol)$, of the unwanted component, $y(\vek{\theta})$, in order to eliminate any residuals of $y(\vek{\theta})$ that would spatially correlate with our component of interest, the temperature-modulated signal $y(\vek{\theta})\, (\Te(\vek{\theta}) - \Teol)$.

The Constrained-ILC method consists in constructing a minimum-variance weighted linear combination of the available frequency maps $\bdx(\vek{\theta})=\{x(\nu,\vek{\theta}) \equiv \Delta I(\nu,\vek{\theta})\}$,
\begin{align}
\label{eq:ilc}
\hat{s}(\vek{\theta}) = \bdw^{\rm T}\, \bdx \equiv \sum_{\nu}\, w(\nu)\, x(\nu,\vek{\theta}),
\end{align}
where $\langle\,\hat{s}^{\,2}\rangle$ is minimized, while the weights $\bdw=\{w(\nu)\}$ are constrained to offer simultaneously unit response to the spectrum ${\partial f(\nu,\Teol) / \partial \Teol}$ of the component of interest, here $y(\vek{\theta})\, (\Te(\vek{\theta}) - \Teol)$, and zero response to the spectrum $f(\nu, \Teol)$ of the unwanted component $y(\vek{\theta})$:
\begin{subequations}
\label{eq:const}
\begin{align}
 \label{eq:const1}
  \bdw^{\rm T}\, \frac{\partial \bdf}{\partial \Teol}= 1,
   \\
  \label{eq:const2}
  \bdw^{\rm T}\, \bdf  = 0.
  \end{align}
\end{subequations}
The spectra $f(\nu, \Teol)$ and ${\partial f(\nu,\Teol) / \partial \Teol}$ are known (Fig.~\ref{Fig:seds}) and can be accurately computed with \textsc{SZpack} \citep{Chluba2012SZpack, Chluba2012moments}.

The optimization problem (Eqs.~\ref{eq:ilc} and \ref{eq:const}) can be solved analytically through Lagrange multipliers, yielding to the so-called Constrained-ILC weights \citep{Remazeilles2011}:
\begin{align}
\label{eq:weights}
   \bdw^{\rm T} &=
  \frac{ \left( \bdf^{\rm T} {\tC}^{-1} \bdf \right)
  \left(\partial_{\Teol} \bdf\right)^{\rm T} {{\tC}}^{-1} - \left( \left(\partial_{\Teol} \bdf\right)^{\rm T}
  {\tC}^{-1} \bdf \right) \bdf^{\rm T}
  {{\tC}}^{-1}
  }{
  \left(\left(\partial_{\Teol} \bdf\right)^{\rm T} {\tC}^{-1} \partial_{\Teol} \bdf
  \right)\left(\bdf^{\rm T} {\tC}^{-1} \bdf \right) -
  \left( \left(\partial_{\Teol} \bdf\right)^{\rm T} {\tC}^{-1} \bdf \right)^2 },
  \end{align}
where ${\tC}_{\rm \nu\nu'} = \langle\,x(\nu)\, x(\nu')\,\rangle$
are the coefficients of the covariance matrix of the data for a pair of frequency maps, and ${\partial_{\Teol} \bdf \equiv {\partial \bdf / \partial \Teol}}$. Therefore, by applying the Constrained-ILC weights Eq.~\eqref{eq:weights} to the data of Eq.~\eqref{eq:model} one obtains that 
\begin{align}
\label{eq:ilc1}
\hat{s}(\vek{\theta}) = y\left(\vek{\theta}\right)\, \left(\Te\left(\vek{\theta}\right) - \Teol\right)\,+\,\bdw^{\rm T}\, \bdn,
\end{align}
which is an unbiased estimate (i.e. no multiplicative bias) of the \textit{temperature-modulated} Compton parameter map ${y(\vek{\theta})\, (\Te(\vek{\theta}) - \Teol)}$, with minimum-variance foreground contamination (term $\bdw^{\rm T}\, \bdn$) and \textit{zero} residuals from the Compton parameter fluctuations $y(\vek{\theta})$ thanks to the constraint Eq.~\eqref{eq:const2}.

By interchanging the constraints Eq.~\eqref{eq:const1} and Eq.~\eqref{eq:const2}, one can alternatively obtain an unbiased estimate of the Compton parameter $y(\vek{\theta})$ with 
\textit{zero} residual contamination
from the temperature-modulated Compton parameter ${y(\vek{\theta})\, (\Te(\vek{\theta}) - \Teol)}$ and minimum-variance foreground contamination:
\begin{align}
\label{eq:ilc2}
\hat{y}(\vek{\theta}) = y\left(\vek{\theta}\right)\,+\,\overline{\bdw}^{\rm T}\, \bdn,
\end{align}
where the new weights $\overline{\bdw}$ are obtained by exchanging $\bdf$ and $\partial_{\Teol} \bdf$ in Eq.~\eqref{eq:weights}. From the clean estimates of $y(\Te - \Teol)$ (Eq.~\ref{eq:ilc1}) and $y$ (Eq.~\ref{eq:ilc2}), having negligible contamination from one into the other, we can easily construct a $y\Te$-map by simply co-adding our estimates as ${\hat{z}\equiv\hat{s} + \Teol\,\hat{y}}$, so that:
\begin{align}
\label{eq:ilc3}
\hat{z}(\vek{\theta}) = y\left(\vek{\theta}\right)\Te\left(\vek{\theta}\right)\,+\,\tilde{\bdw}^{\rm T}\, \bdn,
\end{align}
where $\tilde{\bdw} = \bdw + \Teol\,\overline{\bdw}$. This allows us to directly determine the $y$-weighted temperature and $y^2$-weighted temperature power spectrum from the data.

As a short digression, it is worth noting an equivalent mathematical formulation to the Constrained-ILC, Eq.~\eqref{eq:weights}, by means of the following \textit{standard} ILC:
\begin{align}
\label{eq:equiv-weights}
  \bdw^{\rm T} &= \frac{\bda^{\rm T} {\tC}^{-1}
  }{
   \bda^{\rm T} {\tC}^{-1} \bda},
    \end{align}
in which the newly defined energy spectrum $\bda$ is obtained by Gram-Schmidt orthogonalisation of the initial spectra $\partial_{\Teol} \bdf$ and $\bdf$ of the individual components $y(\Te-\Teol)$ and $y$:
\begin{align}
\label{eq:gram-schmidt}
\bda = \partial_{\Teol} \bdf\, -\, \left( \frac{\left(\partial_{\Teol} \bdf\right)^{\rm T} {\tC}^{-1} \bdf }{\bdf^{\rm T} {\tC}^{-1} \bdf} \right)\,\bdf.
 \end{align}
It can easily be verified that the newly defined \textit{standard} ILC weights Eq.~\eqref{eq:equiv-weights} offer unit response to $y(\Te-\Teol)$ and zero response to $y$ since $\bda^{\rm T} {\tC}^{-1} \bdf = 0$, equivalently to the Constrained-ILC Eq.~\eqref{eq:weights}.

The Constrained-ILC weights, Eq.~\eqref{eq:weights}, can be recast into the compact form:
\begin{align}
\label{eq:compact-weights}
  \bdw^{\rm T} & =  \bde^{\rm T}\,\left({\tA}^{\rm T}\,{\tC}^{-1}\,{\tA}\right)^{-1}\, \tA^{\rm T}\,{\tC}^{-1},
  \end{align}
where ${\tA}=[\partial_{\Teol} \bdf\, \bdf]$ and $\bde=[1\, 0]^{\rm T}$. The Constrained-ILC weights Eq.~\eqref{eq:compact-weights} can then be easily generalised to account for additional constraints complementing Eq.~\eqref{eq:const} in order to null out additional foregrounds. In particular for the present analysis, we wish to null out the kSZ contamination to avoid biases due to spatial correlations between the kSZ signal and $y(\Te-\Teol)$. This also reduces residuals of CMB temperature anisotropies, which have the same spectral shape, $\bda_{\rm CMB, kSZ}(\nu)$, as kSZ.

As already emphasised above, high-frequency channels are essential for breaking temperature degeneracies of the relativistic thermal SZ spectrum at low frequency (Fig.~\ref{Fig:seds}); however, high-frequency channels are also strongly contaminated by Galactic thermal dust emission. We therefore add an additional constraint against the zeroth moment of thermal dust spectrum, ${\bda_{\rm dust}(\nu) \propto \nu^{\overline{\beta}_{\rm d}}\,B(\nu,\overline{T}_{\rm d})}$, where the average dust spectral index is $\overline{\beta}_{\rm d}=1.6$ and the average dust temperature is $\overline{T}_{\rm d}=19.4$\,K \citep{Planck2016GNILC}, and $B(\nu,T)$ is the Planck's blackbody function. This last constraint cannot eliminate completely the residual thermal dust emission because of spatial variations of the dust emissivity and temperature around their mean across the sky, but deprojecting the zeroth moment of the dust in the component separation process can at least remove bulk of the thermal dust foreground contamination. Therefore, we end up with four constraints in our Constrained-ILC approach:
\begin{subequations}
\label{eq:constbis}
\begin{align}
 \label{eq:constbis-1}
  \bdw^{\rm T}\, \frac{\partial \bdf}{ \partial \Teol}= 1,
   \\
   \label{eq:constbis-2}
  \bdw^{\rm T}\, \bdf  = 0,
   \\
   \label{eq:constbis-3}
  \bdw^{\rm T}\, \bda_{\rm CMB,KSZ}  = 0,
   \\
   \label{eq:constbis-4}
  \bdw^{\rm T}\, \bda_{\rm dust}  = 0,
  \end{align}
\end{subequations}
so that the final form of the Constrained-ILC weights in our analysis is generalised to:
\begin{align}
\label{eq:final-weights}
  \bdw^{\rm T} & = \bde^{\rm T}\, \left({\tA}^{\rm T}\,{\tC}^{-1}\,{\tA}\right)^{-1}{\tA}^{\rm T}\,{\tC}^{-1},
  \end{align}
where ${{\tA}=[\partial_{\Teol} \bdf\, \bdf\, \bda_{\rm CMB,KSZ}\, \bda_{\rm dust}]}$ and ${\bde=[1\, 0\, 0 \,0]^{\rm T}}$. We find that this augmentation of the Constrained-ILC method greatly reduces foreground biases in the temperature reconstruction. It can be viewed as a {\it semi-blind method}, since we add partial constraints to the less known dust component.

\subsection{Electron temperature estimation}\label{subsec:temp_est}

From the $y$-map estimate, $\hat{y}(\vek{\theta})$ (Eq.~\ref{eq:ilc2}), and the $y\Te$-map estimate, $\hat{z}(\vek{\theta})$ (Eq.~\ref{eq:ilc3}), we can derive clean estimates of the electron gas temperature $\Te$ of galaxy clusters as follows. A first estimate concerns the \textit{local} temperature profile of individual clusters across the sky, which can be derived from the ratio between the cluster profile of the $y\Te$-map and the cluster profile of the $y$-map:
\begin{align}
\label{eq:te2}
\Tewh^{\,y}(r) = \frac{\hat{z}(r)}{ \hat{y}(r)} \simeq \frac{y(r)\Te(r)}{y(r)},
\end{align}
yielding the $y$-weighted temperature of the cluster, and $r=\vert\vek{\theta} - \vek{\theta}_{\rm c}\vert$ with $\vek{\theta}_{\rm c}$ the coordinate of the centre of the cluster. In parallel, we estimate the electron temperature of galaxy clusters across the sky by computing the mean fluxes of the reconstructed $y\Te$- and $y$-maps within radius $R_{500}$\footnote{$R_{500}$ denotes the radius in which the density of a cluster equals $500$ times the critical density of the Universe. It can be estimated using the $y$-parameter and redshift of the cluster, which we assume are available.} around each cluster, and then taking the ratio:
\begin{align}
\label{eq:te}\widehat{T}^{\,y}_{\rm e, 500} = \frac{\langle\, \hat{z}\,(\vek{\theta})\, \rangle }{ \langle\, \hat{y}\,(\vek{\theta})\, \rangle}\,\Biggr\rvert_{\,\vert\vek{\theta} - \vek{\theta}_{\rm c}\vert\, \leq\, R_{500}} \simeq \frac{\langle\, y\,(\vek{\theta})\,\Te\,(\vek{\theta})\, \rangle }{ \langle\, y\,(\vek{\theta})\, \rangle}\,\Biggr\rvert_{\,\vert\vek{\theta} - \vek{\theta}_{\rm c}\vert\, \leq\, R_{500}}.
\end{align}
This yields the mean $y$-weighted temperature within $R_{500}$ for each galaxy cluster.

A second estimate concerns the \textit{average} temperature over the entire sky across different angular scales, which can be obtained from the cross-power spectrum between the $y$-map and the $y\Te$-map divided by the auto-power spectrum of the $y$-map:
\begin{align}
\label{eq:te1}
\Tewh^{\,yy}(\ell) = \frac{\langle \hat{y}_{\ell m},\hat{z}^*_{\ell m}\rangle }{\langle \vert\hat{y}_{\ell m}\vert^2 \rangle} \simeq \frac{\langle y_{\ell m}, (y\Te)^*_{\ell m}\rangle }{ \langle \vert y_{\ell m}\vert^2 \rangle}  \simeq \frac{C_\ell^{y,y\Te}}{ C_\ell^{yy}}.
\end{align}
This defines the scale-dependent $y^2$-weighted average temperature over the sky and was shown to be the relevant average temperature for unbiased SZ power spectrum analysis \citep{Remazeilles2019}, in contrast to analyses relying on the non-relativistic limit of the thermal SZ effect, which by neglecting temperature corrections underestimate the SZ power spectrum.

Measuring a non-zero temperature would stand for a detection of relativistic SZ effects, as has already been investigated through stacking a sample of massive galaxy clusters \citep{Hurier2016,Erler2018} or through averaging the thermal SZ emission over the entire sky in future spectral distortion measurements \citep{Hill2015}.  We emphasise that our analysis goes beyond a single detection since we aim at mapping and measuring individual cluster temperatures, $\Tewh^{\,y}(\vek{\theta})$, and radial profiles, $\Tewh^{\,y}(r)$, across the sky, as well as the sky-average temperature $\Tewh^{\,yy}(\ell)$ across angular scales.

\subsection{Why is a pivot temperature of $\Teol = 0$ inappropriate?}

At electron temperature $\Te\lesssim {\rm few} \times {\rm keV}$, relativistic temperature corrections can be modeled using an asymptotic expansion of the Compton collision term in orders of $k\Te/\me c^2\ll 1$ \citep{Sazonov1998, Challinor1998,Itoh1998}. This approach is similar to a moment expansion of the thermal SZ frequency spectrum around a pivot temperature of $\Teol = 0$:
\begin{align}
\label{eq:itoh}
\Delta I^{\rm tSZ}\left(\nu,\vek{\theta}\right) &\simeq Y_0\left(\nu\right)\, y\left(\vek{\theta}\right)\, +\, Y_1\left(\nu\right)\, y\left(\vek{\theta}\right)\Te\left(\vek{\theta}\right)+\,\mathcal{O}\left(\Te^2\right),
\end{align}
where the spectral shape $Y_1(\nu)$, e.g., can be found from Eq. (2.27) of \cite{Itoh1998}, while $Y_0(\nu)$ is the non-relativistic spectrum whose analytic expression is given by Eq.~\eqref{eq:non-rel}. 
However, the expansion in Eq.~\eqref{eq:itoh} around $\Teol=0$ does not converge properly for hot clusters with $k\Te\gtrsim 5$\,keV, for which a moment expansion around non-zero temperature (i.e., Eq.~\ref{eq:moment}) would be more reliable. 
This is also illustrated in the top panels of Fig.~\ref{Fig:all-sky-te} and Fig.~\ref{Fig:average_te}, where we show that adopting Eq.~\eqref{eq:itoh}, i.e. a pivot temperature $\Teol=0$ (grey line), in the analysis noticeably underestimates the electron gas temperature of most clusters and the diffuse electron gas temperature in the sky over a broad range of angular scales. 
\footnote{The impact of assuming $\Teol=0$ on the inference of the cosmological parameter $\sigma_8$ has been investigated in \cite{Remazeilles2019}, although this first study is still dependent on the assumed mass bias for the clusters. The impact on the comprehensive set of parameters is still under investigation.}
The cause of this bias is that most clusters have an average temperature of $k\Te\gtrsim 5$\,keV on a broad range of scales, such that higher-order terms in the expansion become significant and can no longer be neglected. This requires estimation of additional \textit{correlated} SZ components (i.e., higher order moments), which if omitted in the component separation process will significantly bias the reconstruction of the $y$ and $\Te$ fields for hot galaxy clusters. 
However, even when adding higher order temperature 
corrections, this problem cannot be overcome for hot clusters, due to the asymptotic convergence of the Taylor-series in $k\Te/\me c^2$ \citep[e.g., Fig.~2 of][]{Chluba2012moments}, such that our approach is deemed to be more robust.

\subsection{Multi-resolution analysis with needlet decomposition}

\begin{figure}
  \begin{center}
    \includegraphics[width=0.97\columnwidth]{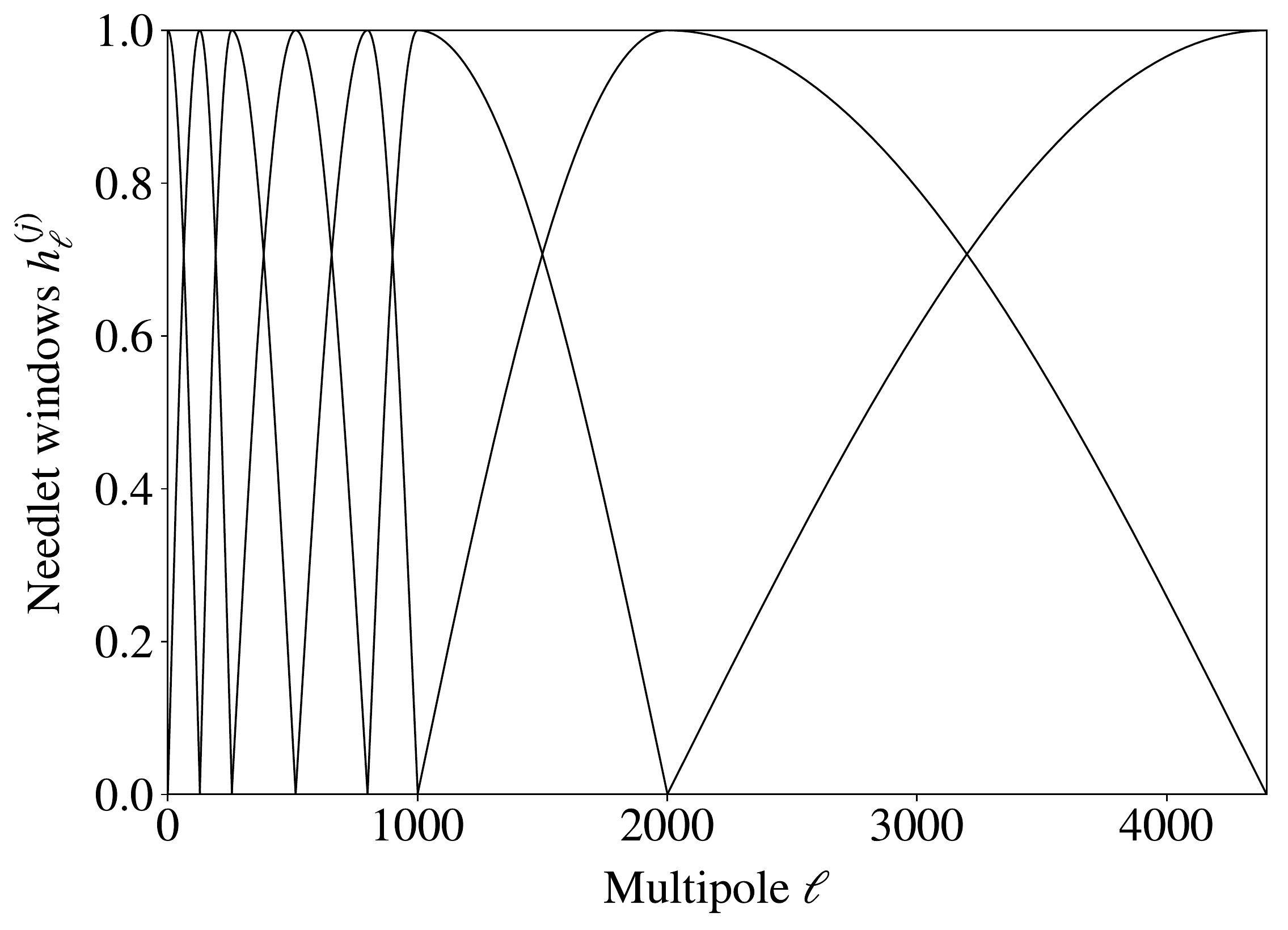}
      \end{center}
\caption{{Needlet windows in harmonic space.} Frequency maps are bandpass-filtered through each needlet window to form a set of needlet maps covering specific ranges of angular scales at each frequency. Afterwards, the multi-frequency component separation analysis is independently performed at each scale.}
\label{Fig:needlets}
\end{figure}

Most component separation methods require to combine multi-frequency maps with the same resolution scale, hence the available data have to be degraded to the lowest channel resolution, which is suboptimal for component separation since it discards the valuable high-multipole correlated information. An interesting alternative is to decompose maps on a \textit{needlet} (spherical wavelet) frame \citep{Narcowich2006,Guilloux2009}, then perform component separation at each wavelet scale independently \citep{Delabrouille2009,Remazeilles2013}. This allows us to combine frequency maps at their native resolution with the Constrained-ILC weights: at large angular scales (first needlet scale), all the frequency channels of an experiment have relevant information so that they can be combined with the Constrained-ILC weights, while towards small angular scales (last needlet scales), low-resolution channel maps add no relevant information, hence they are discarded and only the subset of high-resolution frequency maps, which still have information at these scales, are combined with the Constrained-ILC weights. The multi-scale approach of the needlet-based Constrained-ILC allows us to perform the reconstruction of the $y$ and $\Te$ fields at much higher effective resolution ($5'$ for \pico, $30'$ for \litebird) than the one imposed by the lowest channel resolution of the experiment.

In our analysis, we thus perform a needlet decomposition of the sky maps as follows: the harmonic coefficients, $a_{\ell m}(\nu)$, of each frequency map are bandpass-filtered in harmonic space through needlet windows, $h_\ell^{(j)}$ for $1\leq j \leq 8$, each of them covering a specific range of multipoles (Fig.~\ref{Fig:needlets}). The inverse spherical harmonic transform of the bandpass-filtered coefficients ${\tilde{a}_{\ell m}^{(j)}(\nu) \equiv h_\ell^{(j)}\,a_{\ell m}(\nu)}$ yields to a set of 8 needlet maps at each frequency $\nu$ for each range of angular scales $(j)$, so that each needlet map contains only the temperature fluctuations of a specific range of angular scales. The pixel resolution of each needlet map relates to the needlet scale $(j)$. The Constrained-ILC filtering is then applied in each pixel of the needlet maps instead of the native sky maps, thus performing component separation independently on each range of scales.  

The excellent localization properties of the needlets in both real space (pixels) and harmonic space (multipoles) enable the Constrained-ILC weights to adjust themselves to the varying local conditions of foreground contamination both over the sky and over the angular scales, thus optimizing the component separation. For more details on the use of needlets in component separation, we refer to \cite{Delabrouille2009,Remazeilles2011b,Remazeilles2013,Basak2012}.

\vspace{12mm}
\section{Sky simulations}
\label{sec:simu}

We use the \textsc{PSM} (Planck Sky Model) package \citep{Delabrouille2013} to generate sky simulations across the frequency bands of the NASA's Probe-class CMB space mission concept \pico\ \citep{PICO2019} and JAXA's CMB satellite experiment \litebird\ \citep{Suzuki2018}. The relativistic thermal SZ spectrum, $f(\nu,\Te)$, was computed with \textsc{SZpack} \citep{Chluba2012SZpack, Chluba2012moments}.

\subsection{Thermal SZ Compton-$y$ and temperature maps}

\begin{figure}
  \begin{center}
    \includegraphics[width=0.97\columnwidth]{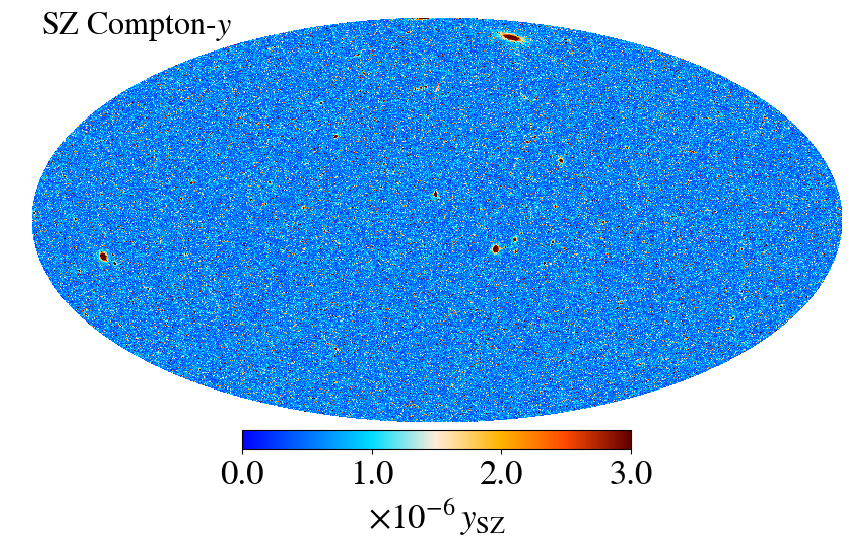}
    \\[2mm]
    \includegraphics[width=0.97\columnwidth]{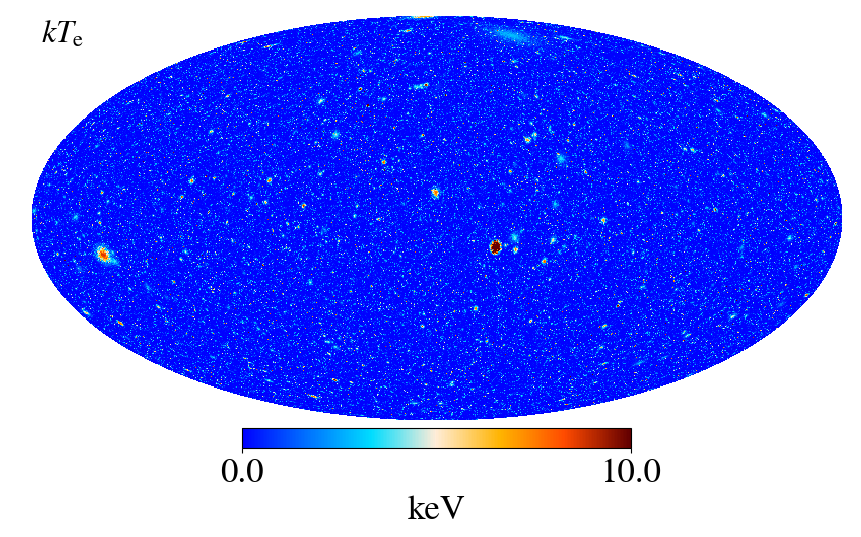}
    \\[2mm]
    \includegraphics[width=0.97\columnwidth]{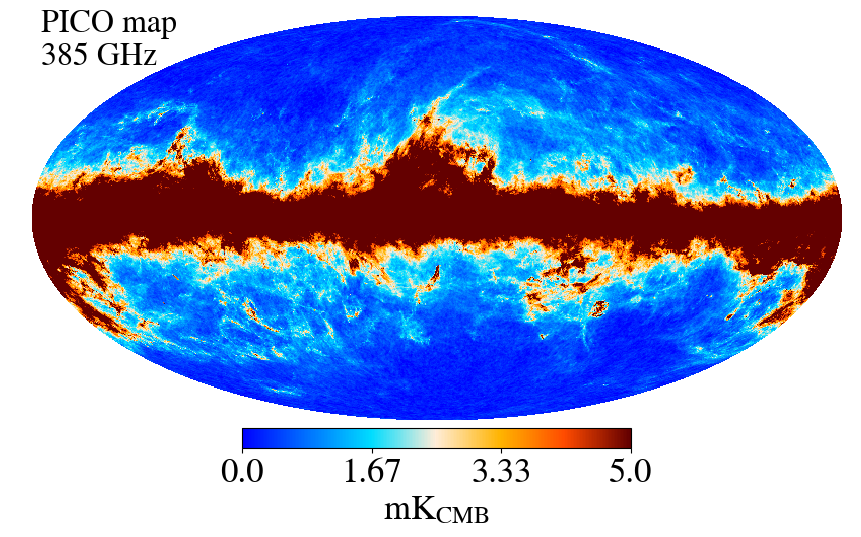}
  \end{center}
\caption{{Sky simulations in Galactic coordinates:} Thermal SZ Compton-$y$ parameter map (\textit{top}); electron temperature field $k\Te$ (\textit{middle}); sky observation map at $385$ GHz for \pico\ (\textit{bottom}), which includes relativistic thermal SZ effects, kinetic SZ effects, CMB, CIB, AME, synchrotron, free-free, thermal dust, and instrumental noise.}
\label{Fig:model}
\end{figure}

The sky maps of the thermal SZ emission from galaxy clusters are built by the \textsc{PSM} by using the position in the sky, flux, size, and temperatures of the galaxy clusters collected in the ROSAT \citep{ROSATclusters} and SDSS \citep{SDSSclusters} cluster catalogues.
The thermal SZ Compton-$y$ parameter map of galaxy clusters is shown in the top panel of Fig.~\ref{Fig:model}. The temperature field $k\Te$ of the relativistic electron gas across the sky is shown in the middle panel of Fig.~\ref{Fig:model}. The electron temperature is spatially correlated with the distribution of galaxy clusters across the sky, as expected.  The dynamical range of electron temperatures across the sky goes from ${k\Te=0}$ to ${k\Te=20}$\,keV in this simulation, with the Coma cluster as an example having an overall temperature\footnote{We found the values for the electron temperatures in the \textsc{PSM} model to be systematically lower that what is expected from simple temperature-mass (T-M) scaling relations \citep[e.g.,][]{Arnaud2005}. 
The origin of this mismatch was due to the introduction of the mass-bias into the T-M scaling relation (while it should enter only in the Y-M relation), which we removed.} 
of $k\Te\simeq 7$\,keV and the Virgo cluster a temperature of $k\Te\simeq 3$\,keV. 

Following Eq.~\eqref{eq:tSZ}, we scale the intensity of the thermal SZ $y$-map across the frequencies using the full relativistic spectrum $f(\nu,\Te)$ computed with \textsc{SZpack}. Thus, the SZ intensity has a different spectral response for each line-of-sight depending on the local temperature $\Te$ across the sky (middle panel of Fig.~\ref{Fig:model}).
Each cluster is mostly modeled as an isothermal sphere, so the input temperature profiles are very close to constant.

\subsection{Foregrounds}
Aside from thermal SZ effects from galaxy clusters, we include several foregrounds in our sky simulations: kinetic SZ effect, CMB, cosmic infrared background (CIB), Galactic synchrotron, anomalous microwave emission (AME), Galactic free-free, and Galactic thermal dust emissions. 

Kinetic SZ effects are due to peculiar radial velocities of galaxy clusters with respect to the CMB rest frame. It constitutes an important foreground to relativistic thermal SZ effects on individual clusters because the velocity profile of the cluster can bias the recovered relativistic temperature profile. CMB temperature anisotropies are also a significant contamination to the observations of galaxy clusters. CIB temperature anisotropies are emitted by dusty star-forming galaxies, which for some of them reside in galaxy clusters, therefore are a correlated foreground to the thermal SZ signal \citep{Addison2012}. 

Aside from extragalactic foregrounds, the largest contribution to the sky emission arises from Galactic foregrounds: synchrotron emission due to high energetic cosmic ray electrons spiralling into Galactic magnetic fields, free-free emission due to free electrons interacting with ionized \textsc{Hii} regions of the Galaxy, thermal dust emission from different populations of dust grains heated by the interstellar medium, and AME arising from electric dipole radiation of ultra-rapid spinning dust grains.

The CMB and kinetic SZ effects are modelled across frequencies through a blackbody spectral shape with $T_{\rm CMB}=2.726$\,K. The CIB emission is simulated in the \textsc{PSM} from three populations of infrared galaxies: spirals, starburst, and proto-spheroid galaxies, as described in \cite{Delabrouille2013}. The Galactic synchrotron component, based on the reprocessed Haslam $408$\,MHz map \citep{Remazeilles2015}, is scaled across frequencies assuming a curved power-law spectrum with a variable spectral index across the sky around a mean of $\beta_{\rm s}\simeq -3$ but with a uniform curvature. The free-free emission is modelled using a power-law with uniform spectral index of $\beta_{\rm ff}=-2.1$. The emission law used to scale AME emission across frequencies is based on the phenomenological CNM model \citep{Ali-Haimoud2009,Draine1998}. For thermal dust emission, the extrapolation across frequencies is performed by assuming a combination of two modified blackbody spectra due to hot and cold dust components with different spectral indices and temperatures \citep{Finkbeiner1999}. The bottom panel of Fig.~\ref{Fig:model} shows the total sky emission map at $385$\,GHz including SZ effects, CMB, CIB, Galactic foregrounds, and instrumental noise from the \pico\ experiment.

\vspace{-3mm}
\subsection{Instruments}
\pico\ \citep{PICO2019} is a NASA's Probe-class CMB satellite mission concept\footnote{\url{https://sites.google.com/umn.edu/picomission/home}} offering unprecedented aggregated sensitivity of $\simeq 0.6\,{\rm \mu K.arcmin}$ in intensity ($\simeq 0.9\,{\rm \mu K.arcmin}$ in polarization), broad frequency coverage (21 frequency bands between 21\,GHz and 799\,GHz), and high resolution. The main instrumental specifications of \pico\ are summarized in Table~\ref{tab:pico}. 

For our mock observation, the thermal SZ and foreground component maps are co-added and integrated over the \pico\ frequency bands, and convolved with Gaussian beams of full-width at half-maximum (FWHM) values listed in Table~\ref{tab:pico}. In addition, white instrumental noise at sensitivity levels quoted in Table~\ref{tab:pico} is added to the sky maps. The set of \pico\ maps thus consists of 21 frequency maps ranging from 21 to 799\,GHz.

Although most of our results are based on the \pico\ simulations, we also consider sky simulations of the CMB space mission \litebird\ \citep{Suzuki2018}, which has been selected by JAXA on 21 May 2019 as one of its next strategic large-class missions\footnote{\url{http://litebird.jp/eng/?p=158}}. \litebird\ has a small telescope aperture compared to \pico\ and therefore rather low angular resolution for galaxy cluster science. Nevertheless, the relatively broad frequency coverage of \litebird\ (15 frequency bands between 40\,GHz and 402\,GHz) and its high sensitivity make it a promising CMB experiment for measuring the relativistic temperature of the largest galaxy clusters like Coma and for constraining the electron temperature power spectrum at the largest angular scales ($\ell \lesssim 200$).

\begin{table}
\caption{Instrumental specifications of the Probe-class CMB satellite project \pico\ \protect\citep{PICO2019} when observing in intensity ($I$).}
  \label{tab:pico}
 \centering
  \begin{tabular}{lll}
\hline\hline
Frequency   & Beam FWHM & Noise $I$ r.m.s \\
$[\rm GHz]$   & $[\rm arcmin]$  & $[\rm \mu K.arcmin]$ \\
\hline 
      21 & 38.4 &   16.9 \\
      25 & 32.0 &   13.0 \\
      30 & 28.3 &   8.8 \\
      36 & 23.6 &   5.6 \\
      43 & 22.2 &    5.6 \\
      52 & 18.4 &    4.0 \\
      62 & 12.8 &    3.8 \\
      75 & 10.7 &    3.0 \\
      90 &  9.5 &    2.0 \\
     108 &  7.9 &    1.6 \\
     129 &  7.4 &    1.5 \\
     155 &  6.2 &    1.3 \\
     186 &  4.3 &    2.8 \\
     223 &  3.6 &    3.2 \\
     268 &  3.2 &    2.2 \\
     321 &  2.6 &   3.0 \\
     385 &  2.5 &   3.2 \\
     462 &  2.1 &  6.4 \\
     555 &  1.5 &  32.4 \\
     666 &  1.3 &  125.2 \\
     799 &  1.1 & 742.5 \\
\hline
   \end{tabular}
\end{table}

\begin{figure}
  \begin{center}
    \includegraphics[width=0.7\columnwidth]{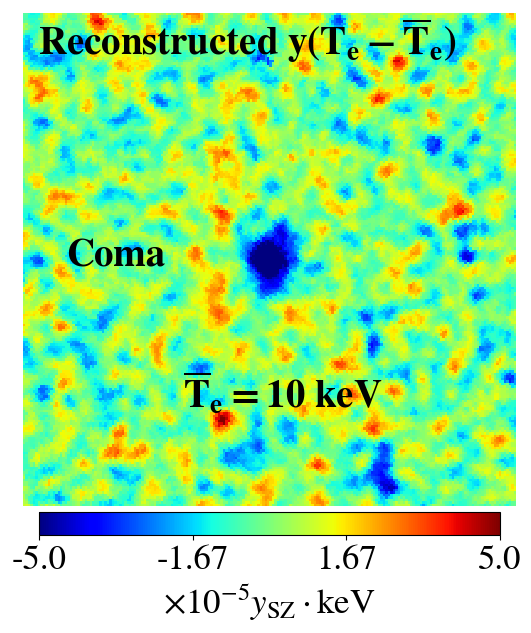}
    \\[0.25mm]
    \includegraphics[width=0.7\columnwidth]{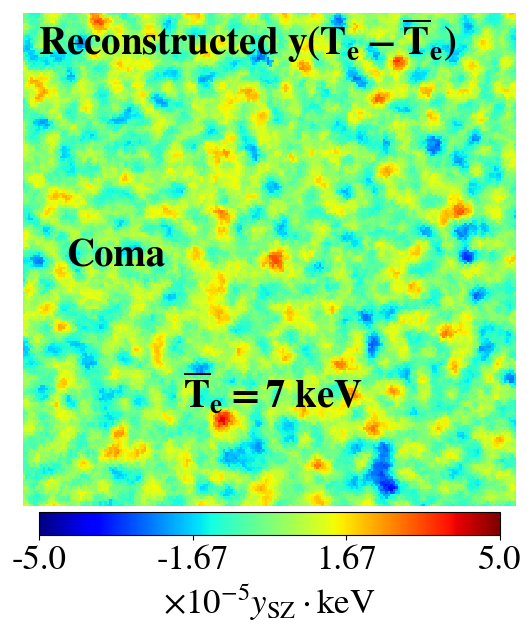}
    \\[0.25mm]
    \includegraphics[width=0.7\columnwidth]{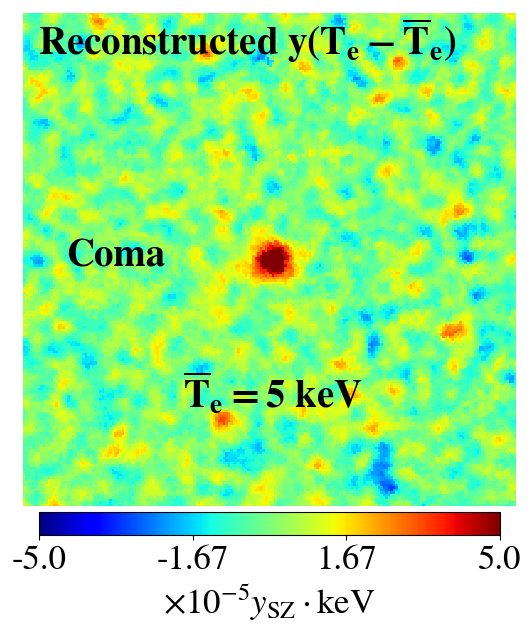}
  \end{center}
\caption{{SZ temperature spectroscopy (Coma):} Reconstruction of the temperature-modulated Compton parameter $y(\Te-\Teol)$-map at $5'$ resolution with \pico\ around Coma cluster ($5^\circ\times 5^\circ$ field of view) with the Constrained-ILC method for different pivot temperatures: ${\Teol=10}$\,keV (\textit{top}); ${\Teol=7}$\,keV (\textit{middle}); ${\Teol=5}$\,keV (\textit{bottom}). This informs us that the temperature of Coma must be closer to ${\Te=7}$\,keV, and that a pivot temperature of ${\Teol=7}$\,keV is the appropriate choice for subsequent analysis.}
\label{Fig:modulated}
\end{figure}
\begin{figure}
  \begin{center}
    \includegraphics[width=0.7\columnwidth]{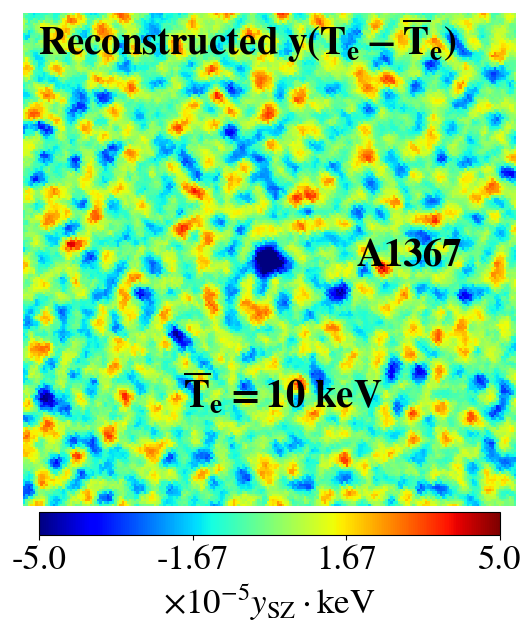}
    \\[0.25mm]
    \includegraphics[width=0.7\columnwidth]{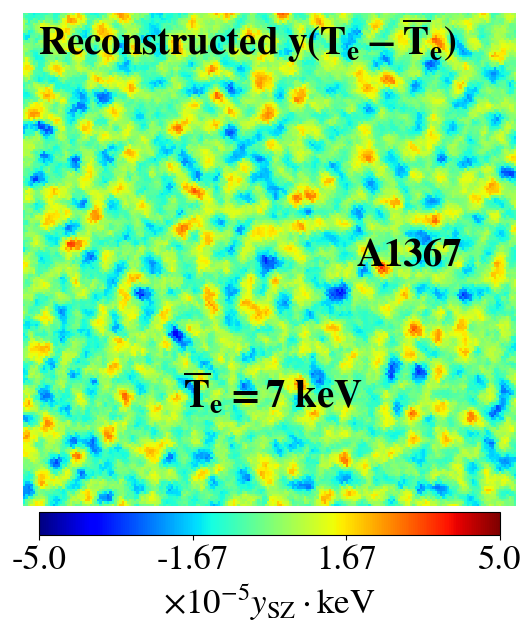}
    \\[0.25mm]
    \includegraphics[width=0.7\columnwidth]{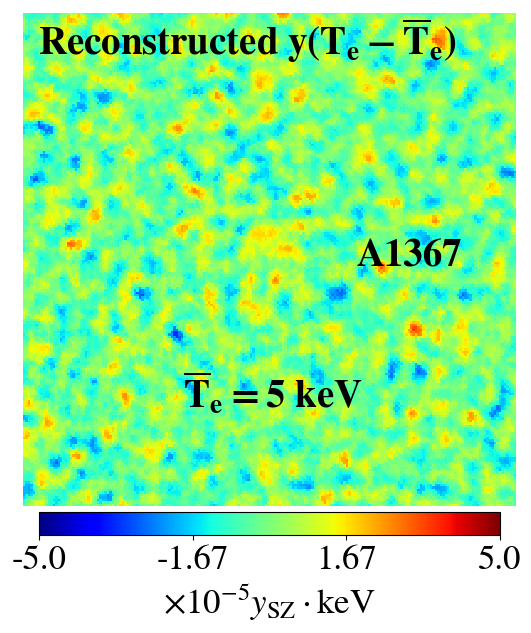}
  \end{center}
\caption{{SZ temperature spectroscopy (Abell~1367):} Reconstruction of the temperature-modulated Compton parameter $y(\Te-\Teol)$-map at $5'$ resolution with \pico\ around Abell~1367 cluster ($5^\circ\times 5^\circ$ field of view) with the Constrained-ILC method for different pivot temperatures: ${\Teol=10}$\,keV (\textit{top}); ${\Teol=7}$\,keV (\textit{middle}); ${\Teol=5}$\,keV (\textit{bottom}). This informs us that for Abell~1367 a pivot temperature of ${\Teol=5}$\,keV is the appropriate choice for subsequent analysis.}
\label{Fig:modulated2}
\end{figure}

\begin{figure*}
  \begin{center}
    \includegraphics[width=\columnwidth]{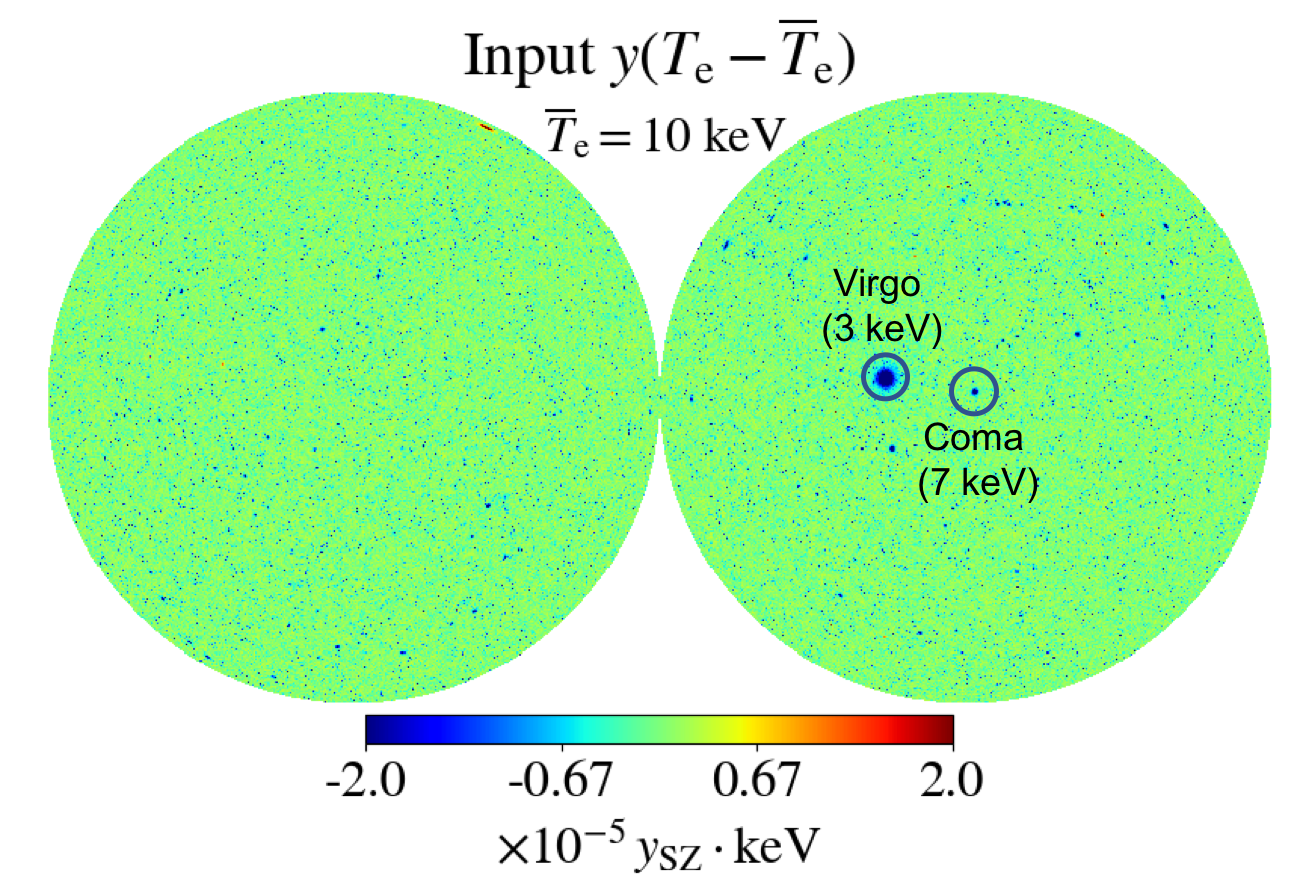}~
    \includegraphics[width=\columnwidth]{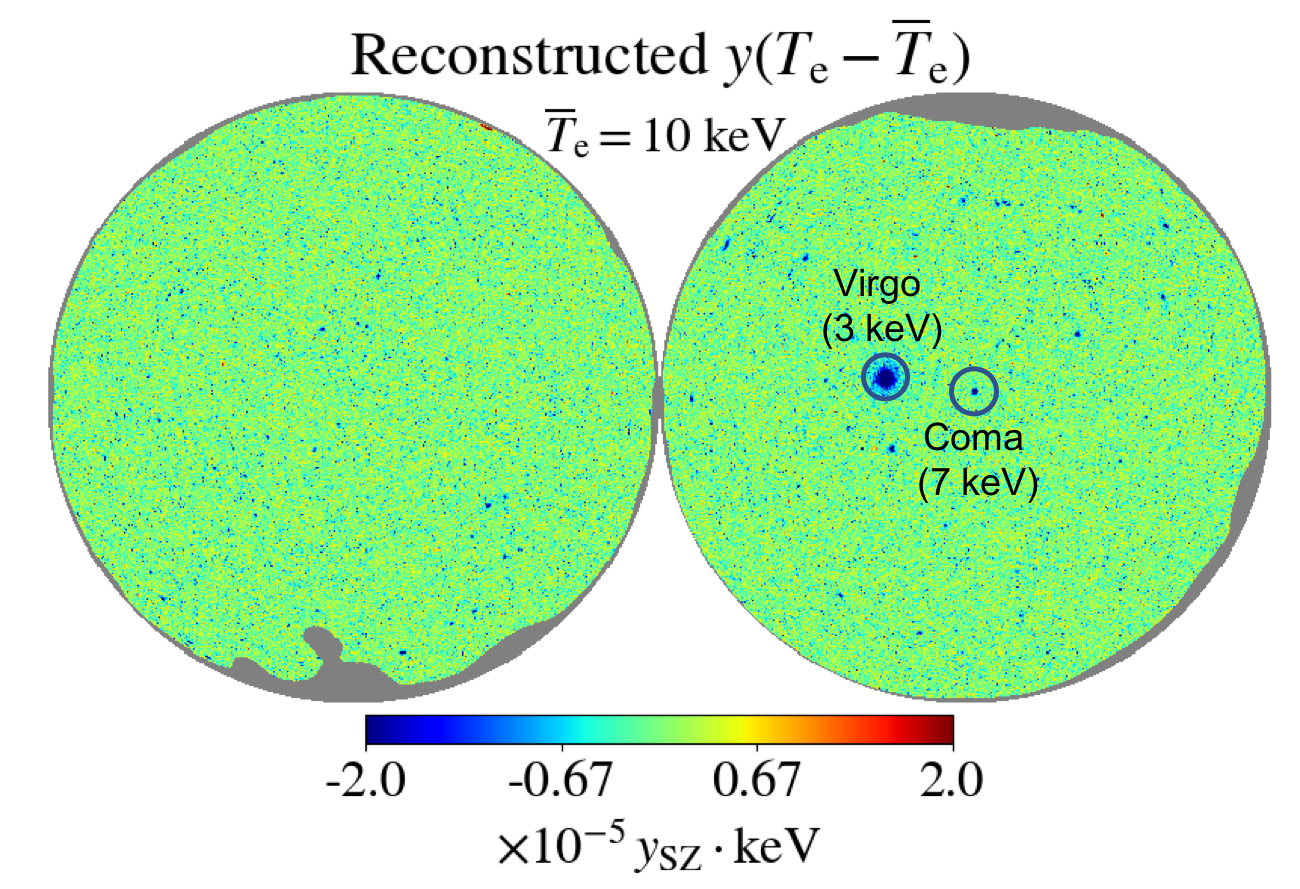}
    \\[2mm]
    \includegraphics[width=\columnwidth]{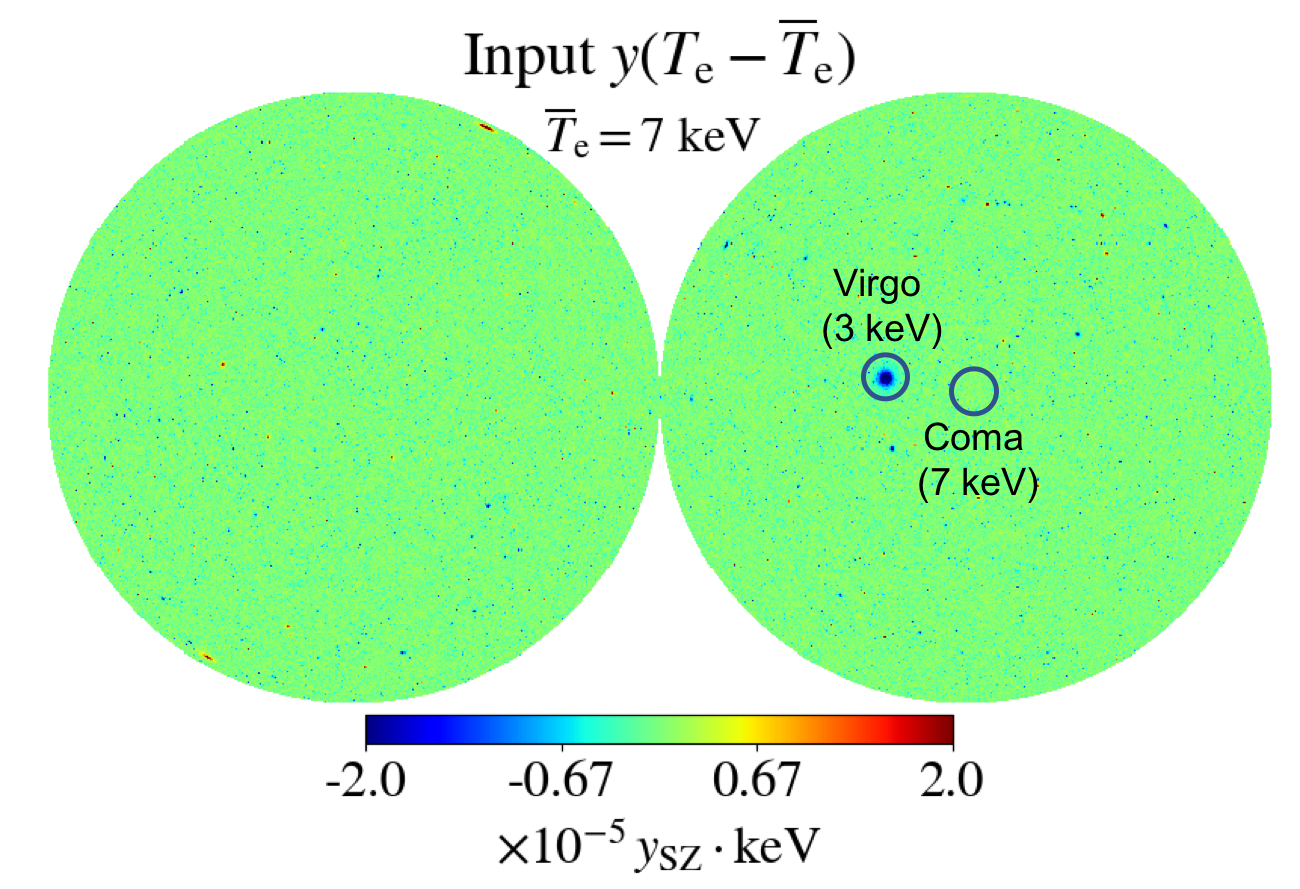}~
    \includegraphics[width=\columnwidth]{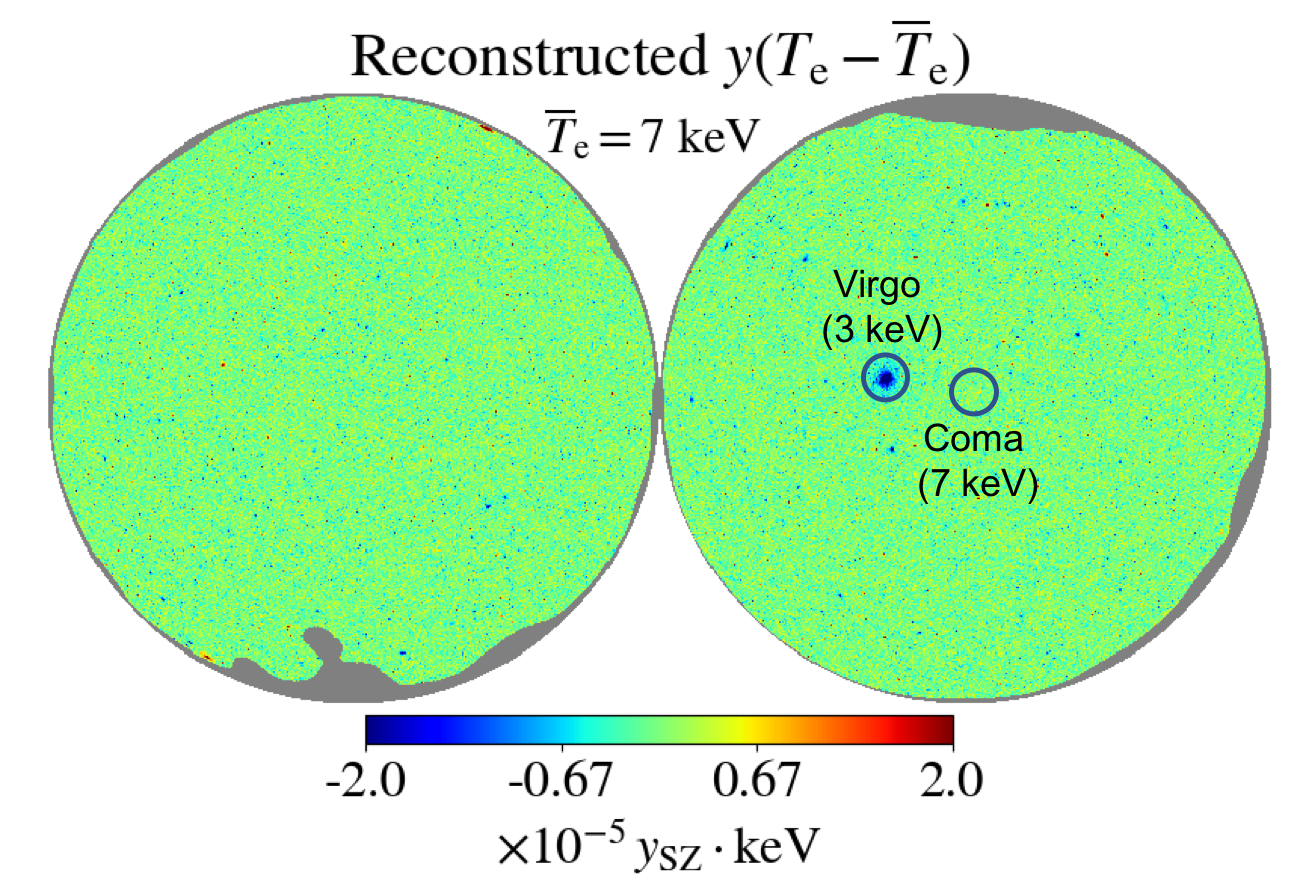}
    \\[2mm]
    \includegraphics[width=\columnwidth]{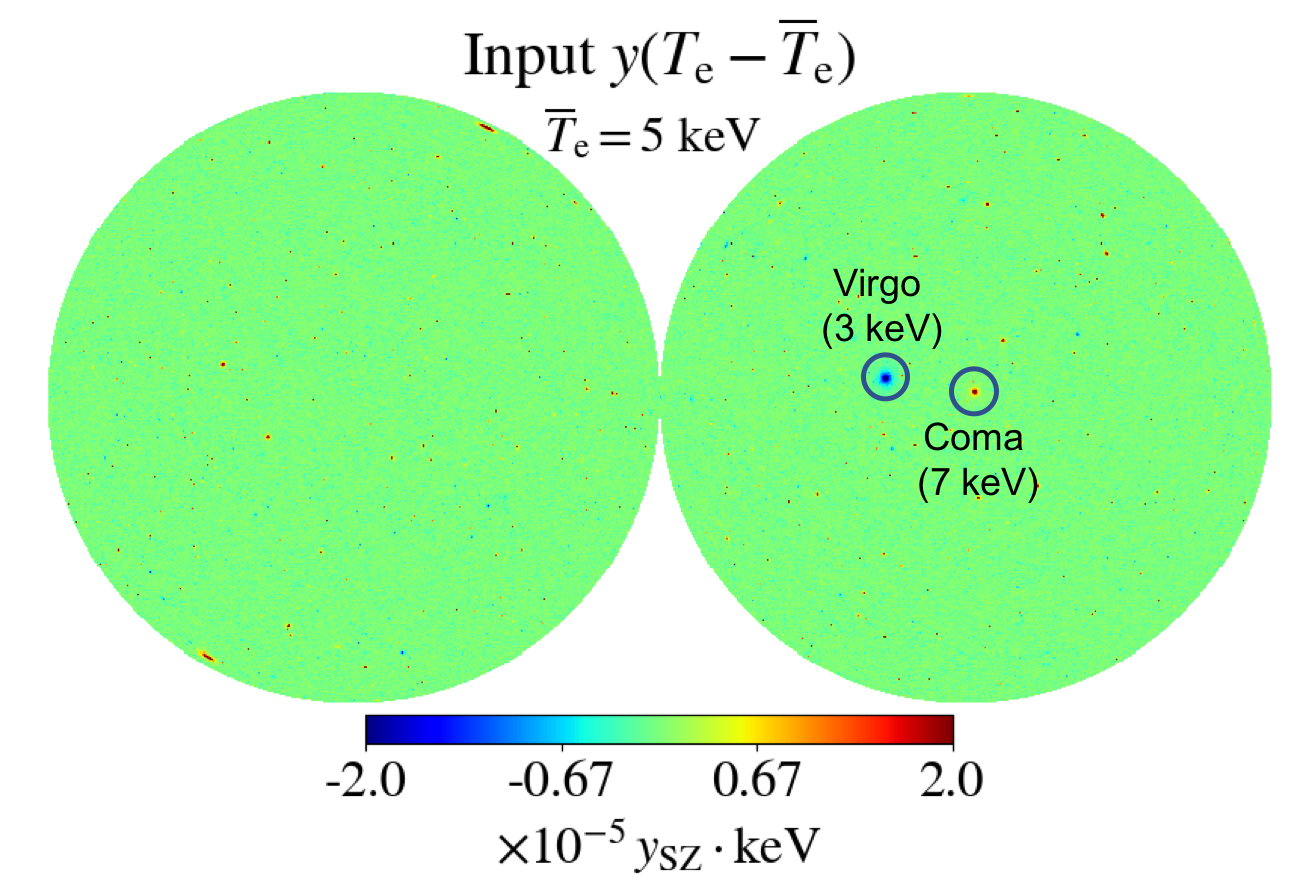}~
    \includegraphics[width=\columnwidth]{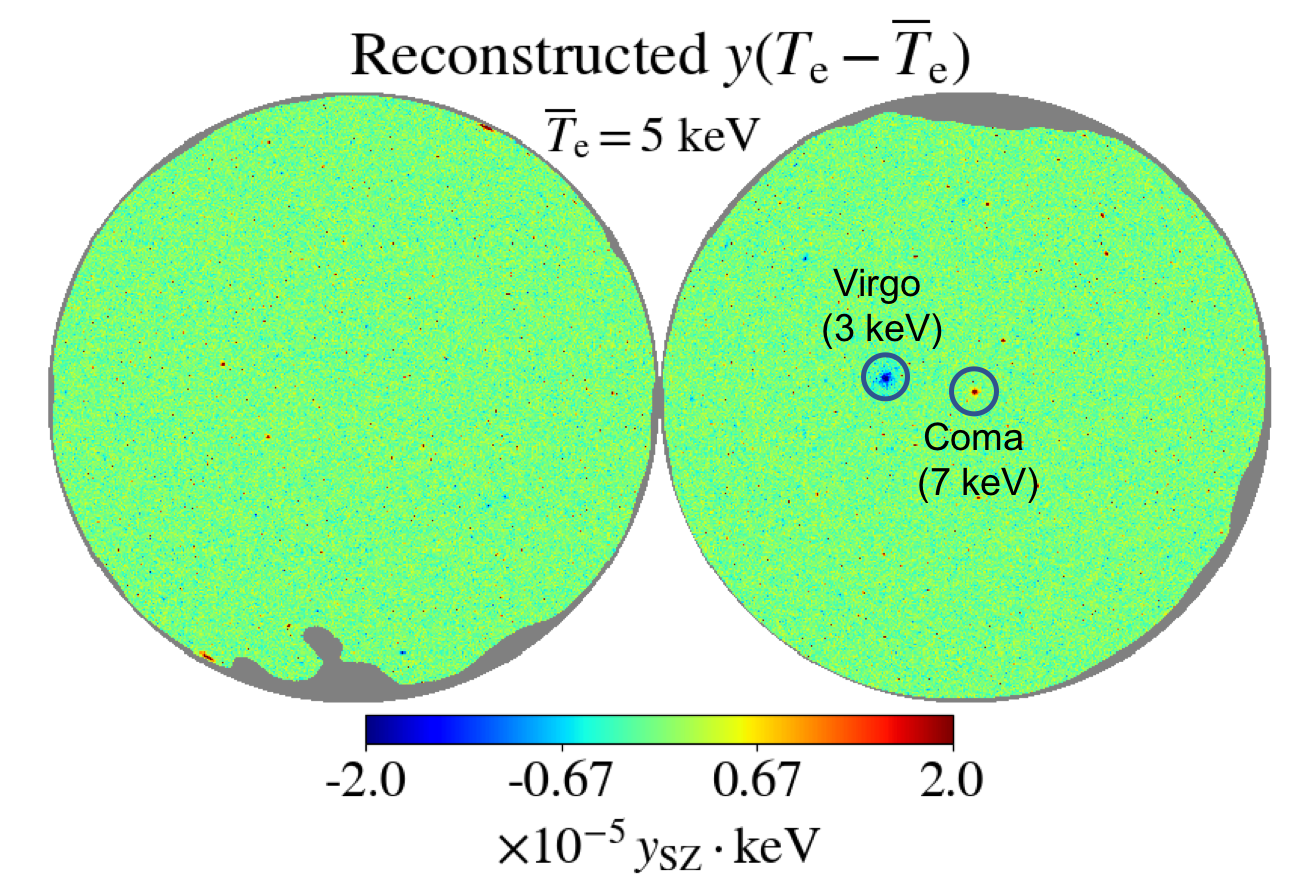}
     \end{center}
\caption{{SZ temperature spectroscopy on the full sky.} \textit{Left}: input temperature-modulated Compton parameter map $y(\Te-\Teol)$ in orthographic projection (left sphere shows southern hemisphere while right sphere shows northern hemisphere in Galactic coordinates) for different pivot temperatures: ${\Teol=10}$\,keV (\textit{top}); ${\Teol=7}$\,keV (\textit{middle}); ${\Teol=5}$\,keV (\textit{bottom}). \textit{Right}: Reconstruction of the temperature-modulated Compton parameter map $y(\Te-\Teol)$ with the Constrained-ILC method for different pivot temperatures: ${\Teol=10}$\,keV (\textit{top}); ${\Teol=7}$\,keV (\textit{middle}); ${\Teol=5}$\,keV (\textit{bottom}). The grey area shows masked pixels in the Galactic plane. For visualisation purposes, the reconstruction is shown without foregrounds (only SZ signal plus noise) and smoothed to $10'$ resolution. Most of the clusters are colder than ${\Teol=10}$\,keV. Coma (centre of northern has a temperature of $\Te\simeq 7$\,keV, while Virgo is colder than $5$\,keV.}
\label{Fig:all-sky}
\end{figure*}

\vspace{-5mm}
\section{Analysis}
\label{sec:analysis}
We apply the Constrained-ILC approach described in Sect.~\ref{subsec:method} to the  \pico\ sky maps in order to clean foregrounds and reconstruct both the $y$-map and the ${y(\Te-\Teol)}$-map, with vanishing contamination from one into the other. We adopt different pivot temperatures of ${\Teol=5}$, $7$, and $10$\,keV\footnote{We  will use on occasion the convention $k=1$, e.g. calling ${\Te=5}$\,keV instead of ${k\Te=5}$\,keV.} in our analysis to perform a "temperature spectroscopy" of the galaxy cluster and determine its actual temperature $\Te$ using map-based tools. 

\vspace{-2mm}
\subsection{Cluster temperature spectroscopy: temperature-modulated ${y(\Te-\Teol)}$-map reconstruction}\label{subsec:spectroscopy}
In Fig.~\ref{Fig:modulated}, we show the reconstruction of the temperature-modulated Compton parameter map ${y(\Te-\Teol)}$ (Eq.~\ref{eq:ilc1}) around the Coma cluster for three different choices of pivot temperatures, ${\Teol=10}$\,keV, ${\Teol=7}$\,keV, and ${\Teol=5}$\,keV, after foreground cleaning.  The effective beam resolution of the reconstructed map is of $5'$. Depending on the pivot temperature adopted in the analysis, the reconstructed component ${\hat{s}=y(\Te-\Teol)}$ (Eq.~\ref{eq:ilc1}) shows either a decrement (negative fluctuation) at the position of the cluster, or an increment (positive fluctuation), or a null. Our component separation approach thus offers a new spectroscopic view of the galaxy clusters, not across frequencies but across temperatures. Clearly, this method enables to take the actual temperature of the Coma cluster after foreground cleaning: it is colder than $10$\, keV (top panel of Fig.~\ref{Fig:modulated}), hotter than $5$\, keV (bottom panel of Fig.~\ref{Fig:modulated}), and close to $7$\,keV (middle panel of Fig.~\ref{Fig:modulated}). In Fig.~\ref{Fig:modulated2}, we show the same kind of temperature-spectroscopic results for a second cluster, Abell~1367, for which the temperature is found to be closer to $5$\,keV (bottom panel of Fig.~\ref{Fig:modulated2}). Our spectroscopic method provides a powerful new tool for accurate SZ analysis of galaxy clusters, using map-based methods.

In addition, we show in Fig.~\ref{Fig:all-sky} the reconstructed temperature-modulated ${y(\Te-\Teol)}$-maps on the \textit{full sky} for different pivot temperatures $\Teol$ versus the input ${y(\Te-\Teol)}$-maps. For clarity, the reconstructed ${y(\Te-\Teol)}$-maps are shown for the case without foregrounds, i.e. only SZ signal plus instrumental noise, and smoothed to 10 arcmin resolution. Despite residual noise, the reconstructed signal on the entire sky matches visually the input signal with high fidelity. Depending on the actual temperature, $\Te$, of the clusters, some of them either vanish in the maps when $\Te\simeq \Teol$, show an increment (red dots) when $\Te > \Teol$, or show a decrement when $\Te < \Teol$. We can appreciate from the distribution of positive and negative clusters in the maps that most of the clusters in the simulated sky are colder than $\Teol=10$\,keV.

\vspace{-2mm}
\subsection{Relativistic $y$-map and $y\Te$-map reconstruction}
Now that we have information on the actual $y$-weighted average temperature of the cluster, $\Te^y\simeq 7$\,keV for Coma or $\Te^y\simeq 5$\,keV for Abell~1367, from our temperature spectroscopy (Fig.~\ref{Fig:modulated}-Fig.~\ref{Fig:modulated2}), we can perform the accurate reconstruction of the relativistic Compton-$y$ parameter from the multi-frequency data by using the appropriate relativistic tSZ energy spectrum, i.e. ${f(\nu,\Teol=7\,{\rm keV})}$ for Coma and ${f(\nu,\Teol=5\,{\rm keV})}$ for Abell~1367, in the component separation process. 

\begin{figure}
  \begin{center}
    \includegraphics[width=0.75\columnwidth]{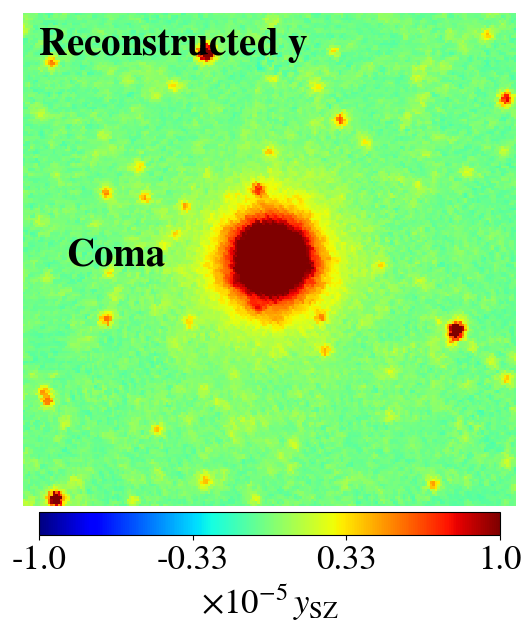}
    \\[2mm]
    \includegraphics[width=0.75\columnwidth]{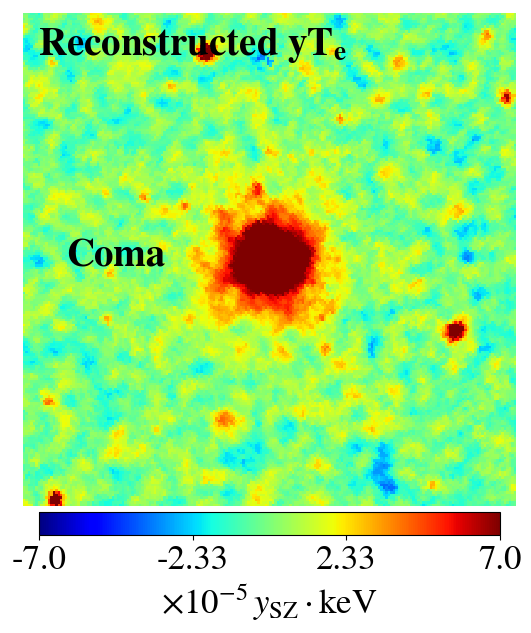}
     \end{center}
\caption{Signal reconstruction at $5'$ resolution with \pico\ around {the Coma cluster} ($5^\circ\times 5^\circ$ field of view) for a pivot temperature of ${\Teol=7}$\,keV: Constrained-ILC $y$-map (\textit{top});  Constrained-ILC $y\Te$-map (\textit{bottom}). We are able to recover the temperature $\Te\simeq 7$\,keV of the Coma cluster through the apparent scaling of the reconstructed $y\Te$-map with respect to the reconstructed $y$-map.}
\label{Fig:maps}
\end{figure}
\begin{figure}
  \begin{center}
    \includegraphics[width=0.75\columnwidth]{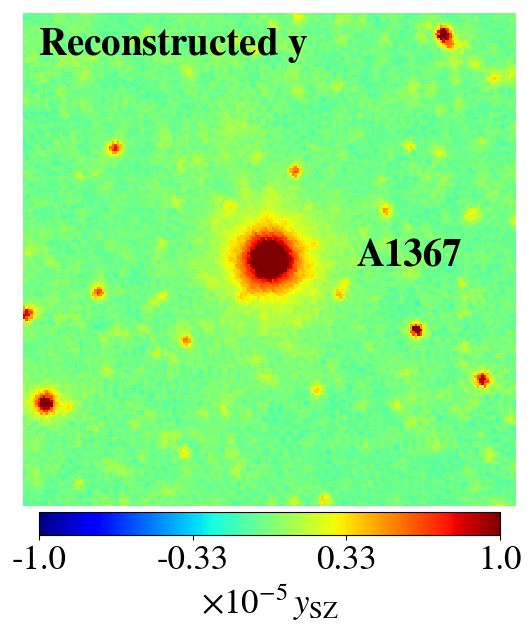}
    \\[2mm]
    \includegraphics[width=0.75\columnwidth]{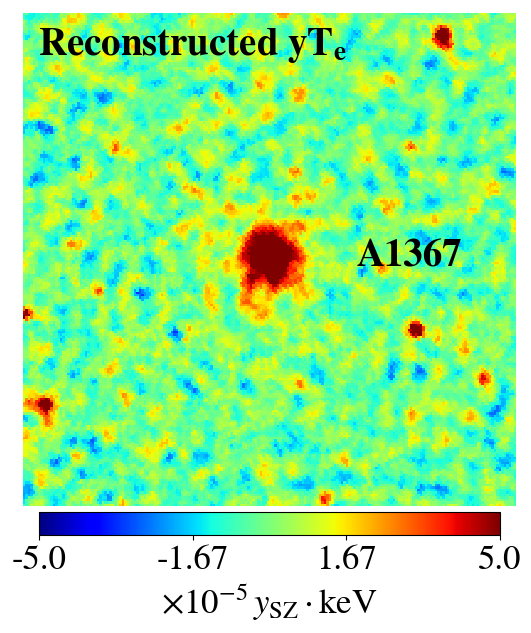}
      \end{center}
\caption{Signal reconstruction at $5'$ resolution with \pico\ around {Abell~1367} ($5^\circ\times 5^\circ$ field of view) for a pivot temperature of ${\Teol=5}$\,keV: Constrained-ILC $y$-map (\textit{top}); Constrained-ILC $y\Te$-map (\textit{bottom}). We are able to recover the temperature $\Te\simeq 5$\,keV of the Abell~1367 cluster through the apparent scaling of the reconstructed $y\Te$-map with respect to the reconstructed $y$-map.}
\label{Fig:maps2}
\end{figure}

The reconstructed relativistic $y$-map at $5'$ resolution obtained for a pivot temperature of $\Teol=7$\,keV and centred at the position of Coma is shown in the upper panel of Fig.~\ref{Fig:maps}. Similarly, the reconstructed $y$-map obtained for a pivot temperature of $\Teol=5$\,keV and centred at the position of Abell~1367 is shown in the upper panel of Fig.~\ref{Fig:maps2}. In both cases, the reconstruction is extremely clean in terms of residual foreground contamination thanks to wide distribution of frequencies and high sensitivity level of the \pico\ experiment.

\begin{figure}
  \begin{center}
    \includegraphics[width=0.97\columnwidth]{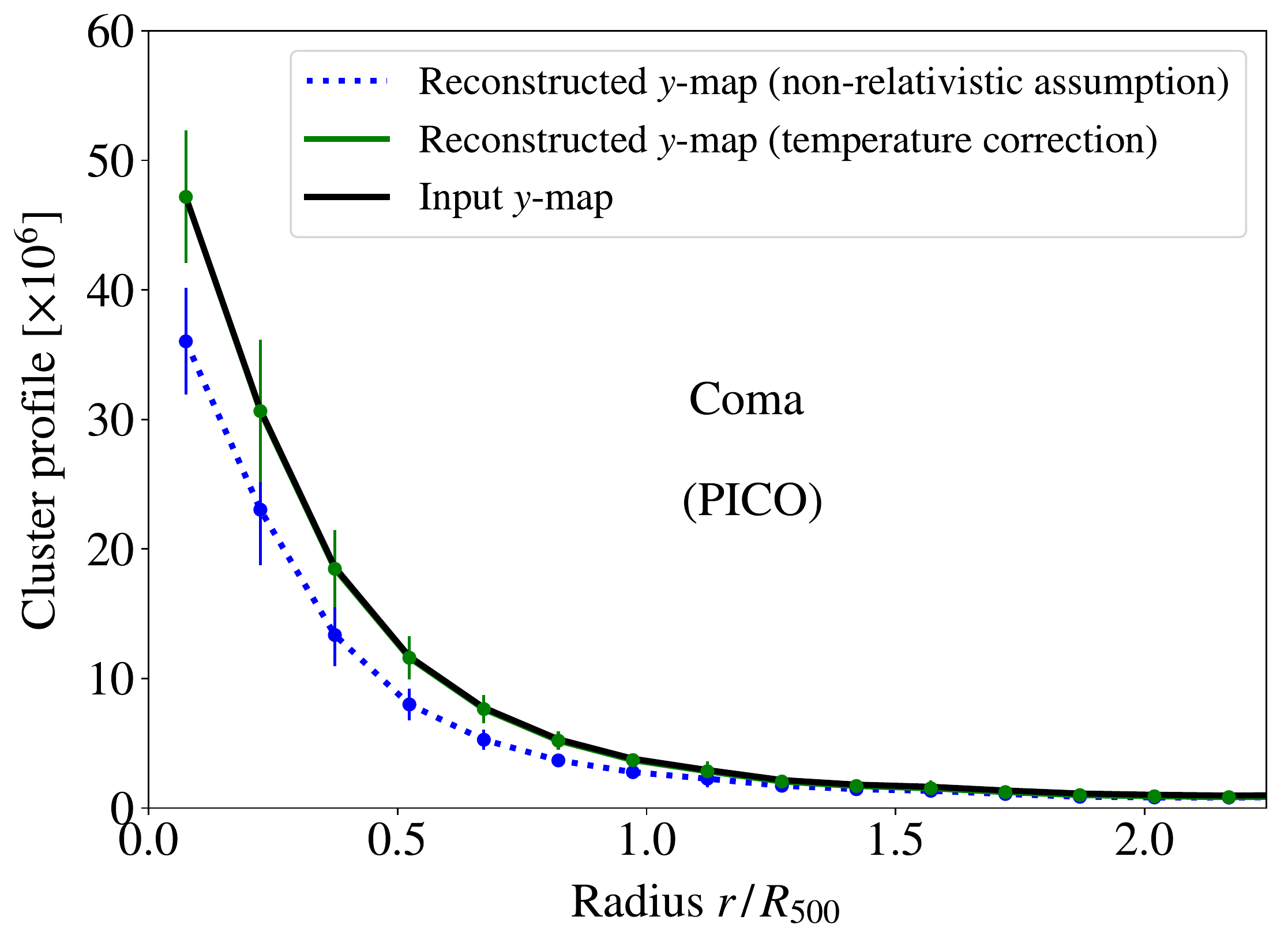}
      \end{center}
\caption{{ Bias in the Compton $y$-profile measurement with \pico\ when neglecting relativistic SZ temperature corrections:} input Coma cluster profile (\textit{solid black}); reconstructed Coma profile with non-relativistic SZ spectrum assumption (\textit{dotted blue}); reconstructed Coma profile when accounting for relativistic temperature correction (\textit{solid green}).}
\label{Fig:y_nonrel_vs_rel}
\end{figure}

In most SZ data analyses to date it is usual practice to adopt the non-relativistic limit to model the tSZ spectrum (i.e., to only include the leading order term $\propto Y_0$) in the component separation pipeline \citep[e.g][]{Planck_int_resultsV2013,Planck2015ymap}. 
In Fig.~\ref{Fig:y_nonrel_vs_rel}, we illustrate the limitations of this practice by computing the reconstructed Coma cluster $y$-profile when either neglecting (non-relativistic limit; dotted blue) or accounting for the temperature of Coma (solid green) in the thermal SZ spectrum, and compare it with the input cluster profile (solid black). Had we neglected relativistic temperature corrections, then the reconstruction of the Compton-$y$ parameter would have been significantly biased and the amplitude of the cluster pressure profile would have been underestimated. Given that the temperature of galaxy clusters depends on their mass and redshift, neglecting relativistic SZ effects will impact both the amplitude and shape of the \textit{mean} pressure profile of galaxy clusters, a quantity on which cosmological analyses of galaxy clusters strongly rely \citep{Ruppin2019}.
 While for individual clusters the bias due to neglecting relativistic temperature corrections is not as relevant at the sensitivity levels of the \Planck\ experiment, it can no longer be ignored for future CMB experiments like \pico\ (Fig.~\ref{Fig:y_nonrel_vs_rel}). Still, even for \Planck\ the bias exists and becomes significant when considering ensembles of clusters (i.e., in stacking, SZ power spectrum analysis, and cluster number count analyses). As discussed in \cite{Remazeilles2019}, this may be part of the source of the \Planck's tension on the cosmological parameter $\sigma_8$ and thus should be considered carefully.

Along with the reconstructed $y$-map (Eq.~\ref{eq:ilc2}), our component separation method allows to reconstruct the $y\Te$-map (Eq.~\ref{eq:ilc3}), with vanishing contamination from one into the other. The reconstructed $y\Te$-map at $5'$ resolution obtained for a pivot temperature of $\Teol=7$\,keV and centred at the position of Coma is shown in the lower panel of Fig.~\ref{Fig:maps}, while the reconstructed $y\Te$-map for a pivot temperature of $\Teol=5$\,keV and centred at the position of Abell~1367 is shown in the lower panel of Fig.~\ref{Fig:maps2}. The reconstruction is noisier than for the $y$-map because the intensity of the $y\Te$ component is obviously fainter. From these first map visualisations, we see that the scaling of the amplitude of the reconstructed $y\Te$ map with respect to the amplitude of the reconstructed $y$-map is driven by the actual temperature of the cluster, i.e., $\Te\simeq 7$\,keV for Coma and $\Te \simeq 5$\,keV for Abell~1367. Therefore, the ratio of the two signals at the cluster positions allows us to derive the actual temperature of the clusters, as we discuss next.

\subsection{Electron temperature $\Te$ reconstruction across the sky}\label{subsec:Te}

From the accurate separation of the Compton-$y$ (upper panels of Figs.~\ref{Fig:maps}-\ref{Fig:maps2}) and \textit{temperature-modulated} Compton-$y\Te$ components (lower panels of Figs.~\ref{Fig:maps}-\ref{Fig:maps2}), several combinations can be performed to infer the electron temperature $\Te^y(\vek{\theta})$ across the full sky, as well as the temperature profiles $\Te^y(r)$ of individual clusters and the average temperature $\Te^{yy}(\ell)$ over the sky across different angular scales, as described in Sect.~\ref{subsec:temp_est} and detailed here.

\begin{figure}
  \begin{center}
       \includegraphics[width=\columnwidth]{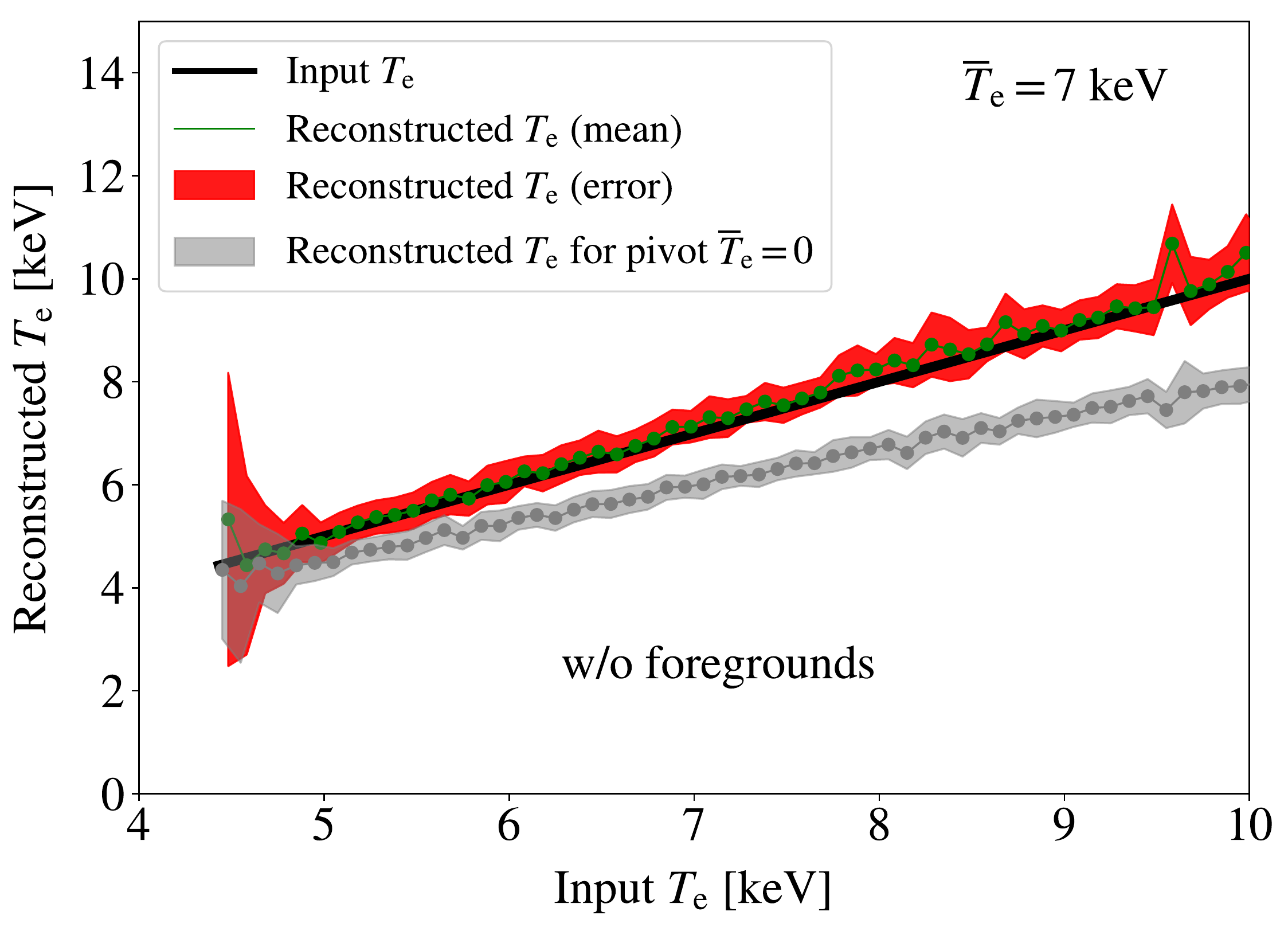}
    \\[2mm]
    \includegraphics[width=\columnwidth]{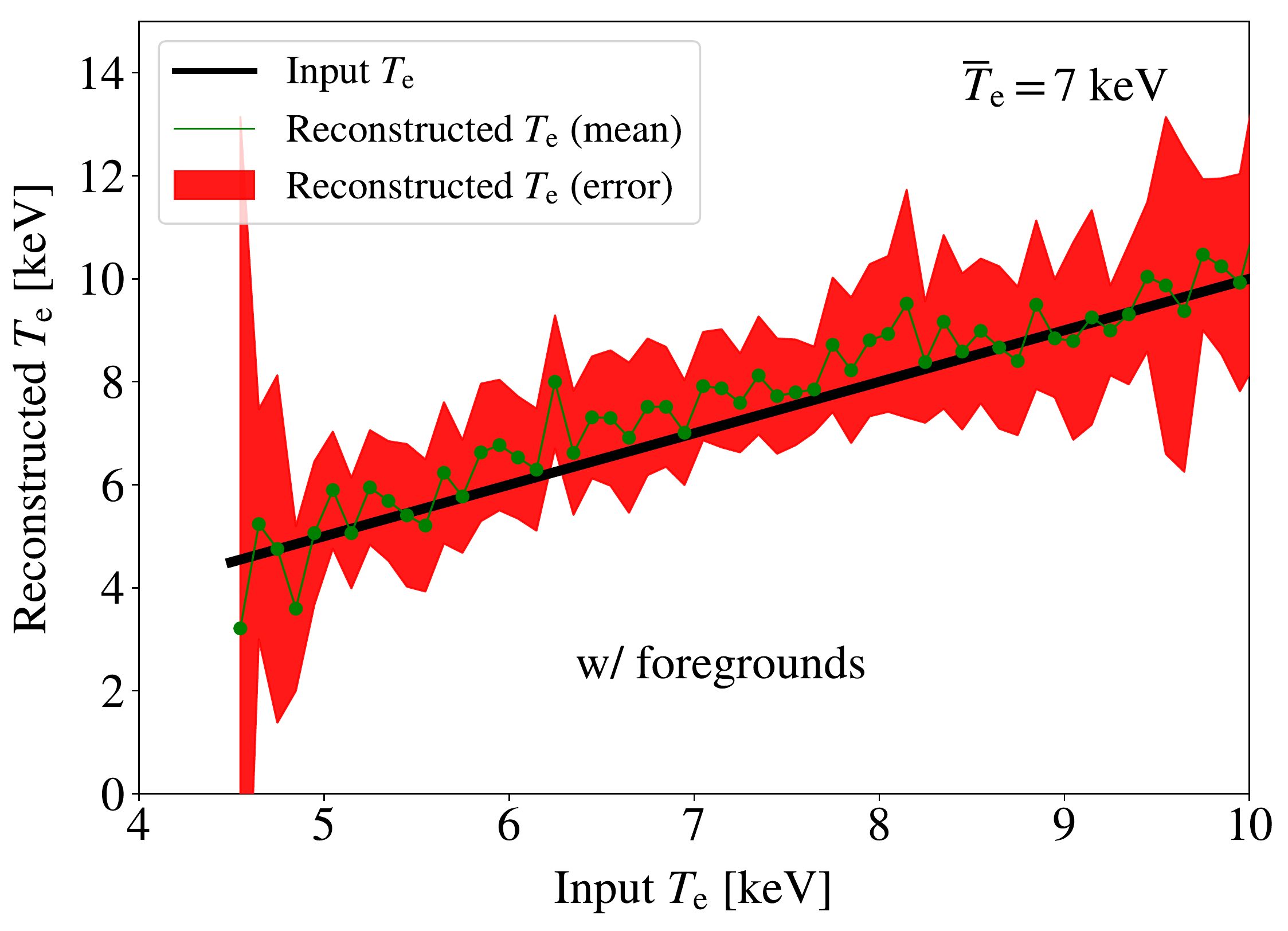}
  \end{center}
\caption{Recovered electron temperatures across the full sky with \pico\ of a sample of 811 clusters having temperatures ranging between $4.5$ and $10$\,keV, for regular bins of $0.1$\,keV. \textit{Upper panel}: without foregrounds. \textit{Lower panel}: with foregrounds.}
\label{Fig:all-sky-te}
\end{figure}

\subsubsection{$\Te$ mapping across the full sky}\label{subsubsec:Te-mapping}

First, from the reconstructed $y$- and $y\Te$ full-sky maps we select the sample of 811 galaxy clusters in our simulation whose temperatures range from $\Te=4.5$\,keV to $\Te=10$\,keV and whose the size is larger than the $5'$ beam resolution of the maps. For each cluster, we compute the average flux of both the $y$- and $y\Te$ signals within an aperture of radius $R_{500}$ and take the ratio of fluxes between the two signals to estimate the actual temperature $\widehat{T}^{\,y}_{\rm e, 500}$ (Eq.~\ref{eq:te}) of each cluster. The uncertainty on $y$ and $y\Te$ for each cluster is computed from the standard deviation of the signals within the aperture. Then, the error bars on the recovered temperatures are computed from the variance of a ratio between two random variables, $\Tewh = \hat{z}(r) / \hat{y}(r)$, which relates to the means $\{\mu_{\hat{z}},\mu_{\hat{y}}\}$ , variances $\{\sigma^2_{\hat{z}},\sigma^2_{\hat{y}}\}$, and covariance ${\rm Cov}(\hat{z},\hat{y})$ of the $y\Te$- and $y$-fluxes:
\begin{align}
\label{eq:stddev_ratio}
\sigma^2(\Tewh) = \frac{(\mu_{\hat{z}})^2 }{ (\mu_{\hat{y}})^2} \left[  \frac{\sigma^2_{\hat{z}} }{ (\mu_{\hat{z}})^2} - 2\frac{{\rm Cov}\left(\hat{z},\hat{y}\right)  }{ \mu_{\hat{z}}\,\mu_{\hat{y}} } + \frac{\sigma^2_{\hat{y}} }{ (\mu_{\hat{y}})^2} \right].
\end{align}

Figure.~\ref{Fig:all-sky-te} shows the recovered temperatures for the large sample of clusters distributed across the full sky and averaged in temperature bins of $\Delta \Te=0.1$\,keV, both in the absence (\textit{upper panel}) and in the presence (\textit{lower panel}) of foregrounds, when adopting a pivot temperature $\Teol=7$\,keV in the component separation analysis. The black line shows the input temperatures of the clusters in the simulation, while the green line shows the reconstructed temperatures after component separation, with error bars indicated by the shaded red area.

Clearly, our component separation approach enables to recover and measure cluster temperatures pretty accurately across the full sky after foreground removal, although the presence of residual foreground contamination increases the uncertainties in the lower panel of Fig.~\ref{Fig:all-sky-te}. A few outliers in temperature recovery arise from clusters in regions of the sky suffering from larger residual foreground contamination, but are still consistent with the input temperature within uncertainties.
As a comparison, we also show in the upper panel Fig.~\ref{Fig:all-sky-te} how the recovered temperatures would be significantly biased (grey line) if a pivot temperature of $\Teol=0$ (Eq.~\ref{eq:itoh}) had been used in the analysis. 
We find the bias in the recovered temperature to scale as ${\Te/\Te^{\rm true} \approx 1-0.15 \,[k\Te/7\,{\rm keV}]}$, implying an underestimation of the cluster temperature by $\gtrsim 10$\% for hot systems.
This highlights the importance of a refined modeling of the relativistic SZ effects in the analysis (Eq.~\ref{eq:moment}) and also emphasises the relevance of a temperature spectroscopy step (Sect.~\ref{subsec:spectroscopy} and Fig.~\ref{Fig:modulated}) in our component separation approach to obtain the appropriate pivot temperature prior to the final iteration in analysis.
By allowing the reconstruction of SZ temperatures across the sky (Fig.~\ref{Fig:all-sky-te}), our method thus provides a new proxy for measuring cluster masses that is independent of X-ray scaling relations \citep[e.g.][]{Arnaud2005,Reichert2011}.

\begin{figure}
  \begin{center}
    \includegraphics[width=0.97\columnwidth]{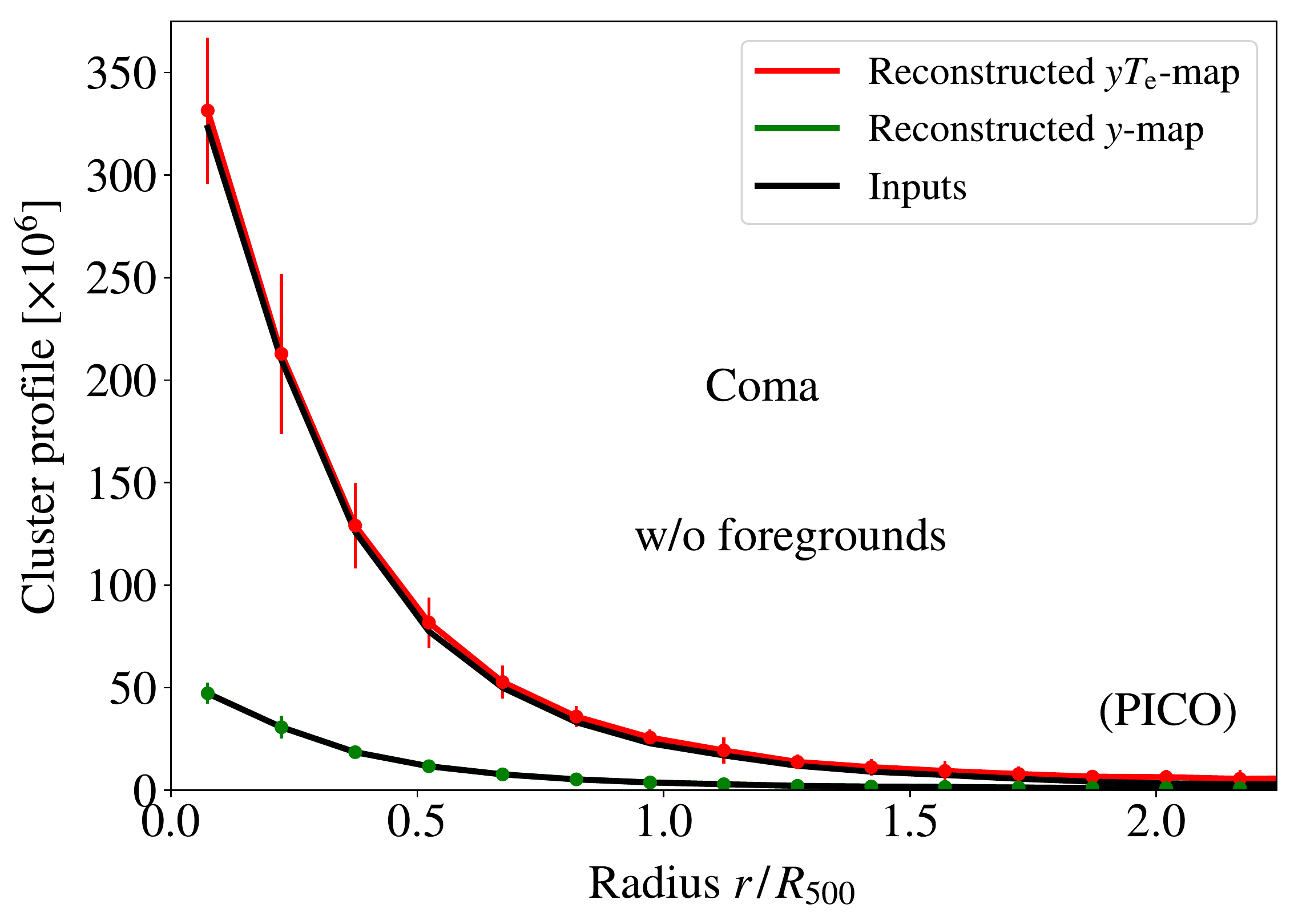}
    \\[2mm]
    \includegraphics[width=0.97\columnwidth]{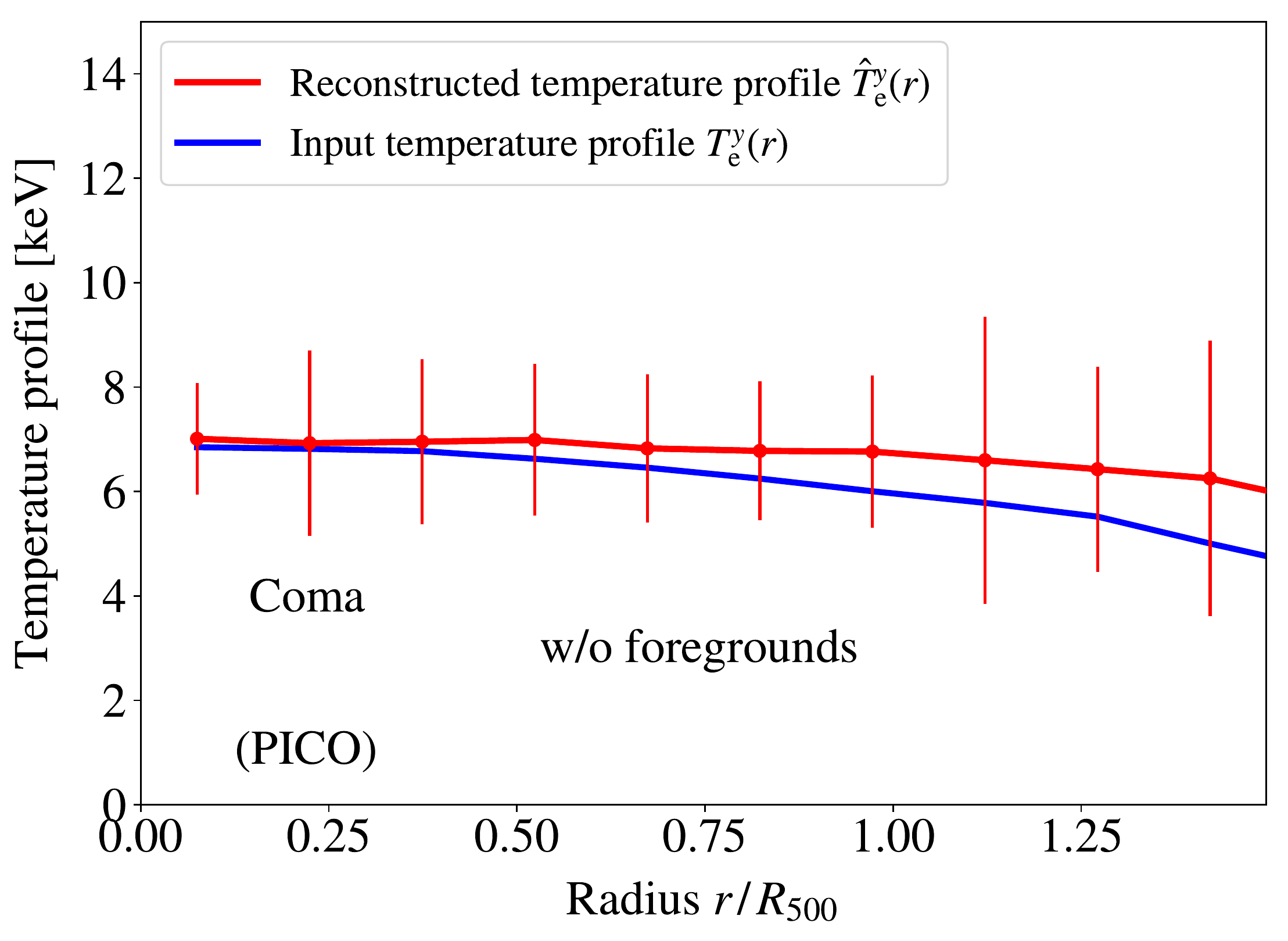}
   \end{center}
\caption{{ \pico\ without foregrounds.} \textit{Top}: Coma cluster profiles from the reconstructed $y$-map (\textit{green}) and the reconstructed $y\Te$-map (\textit{red}). \textit{Bottom}: Reconstructed temperature profile of Coma cluster from the ratio of the $y\Te$ profile and the $y$ profile (Eq.~\ref{eq:te2}).}
\label{Fig:profiles_and_power_nofg}
\end{figure}

\begin{figure}
  \begin{center}
    \includegraphics[width=0.97\columnwidth]{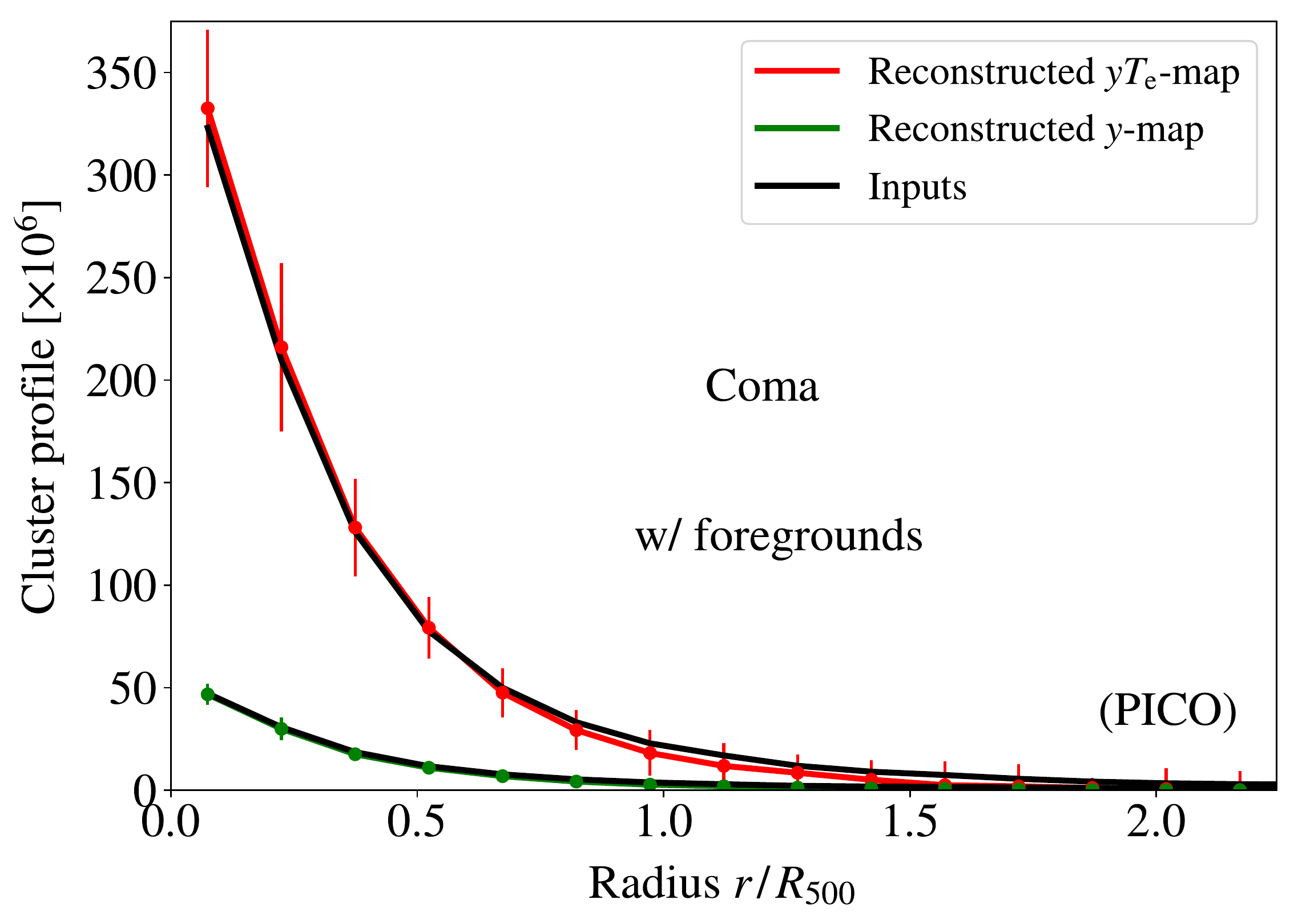}
    \\[2mm]
    \includegraphics[width=0.97\columnwidth]{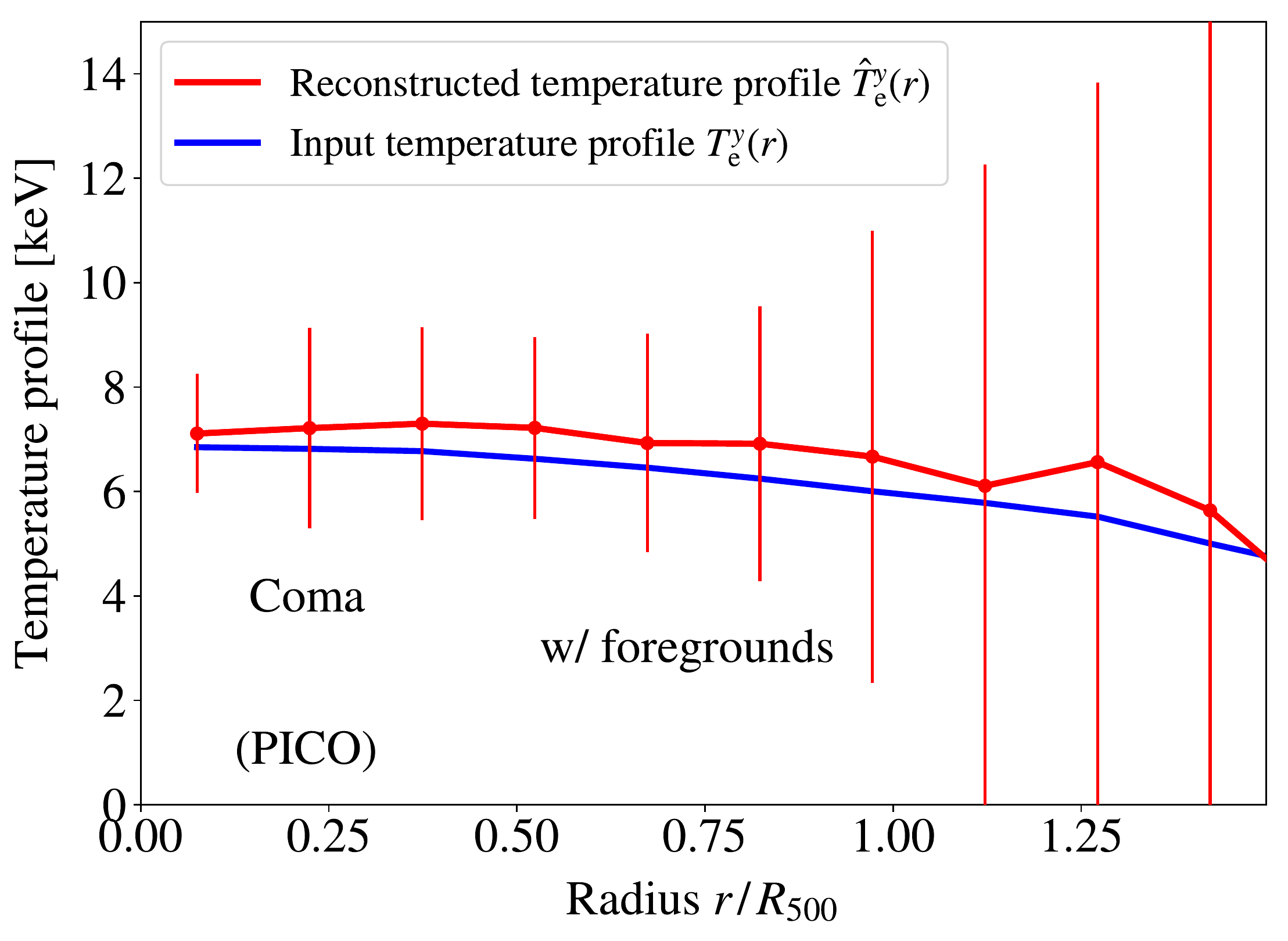}
  \end{center}
\caption{{ \pico\ with foregrounds.} \textit{Top}: Coma cluster profiles from the reconstructed $y$-map (\textit{green}) and the reconstructed $y\Te$-map (\textit{red}). \textit{Bottom}: Reconstructed temperature profile of Coma cluster from the ratio of the $y\Te$ profile and the $y$ profile (Eq.~\ref{eq:te2}).}
\label{Fig:profiles_and_power}
\end{figure}

\subsubsection{$\Te$ profiles}\label{subsubsec:Te-profile}
\begin{figure}
  \begin{center}
    \includegraphics[width=0.97\columnwidth]{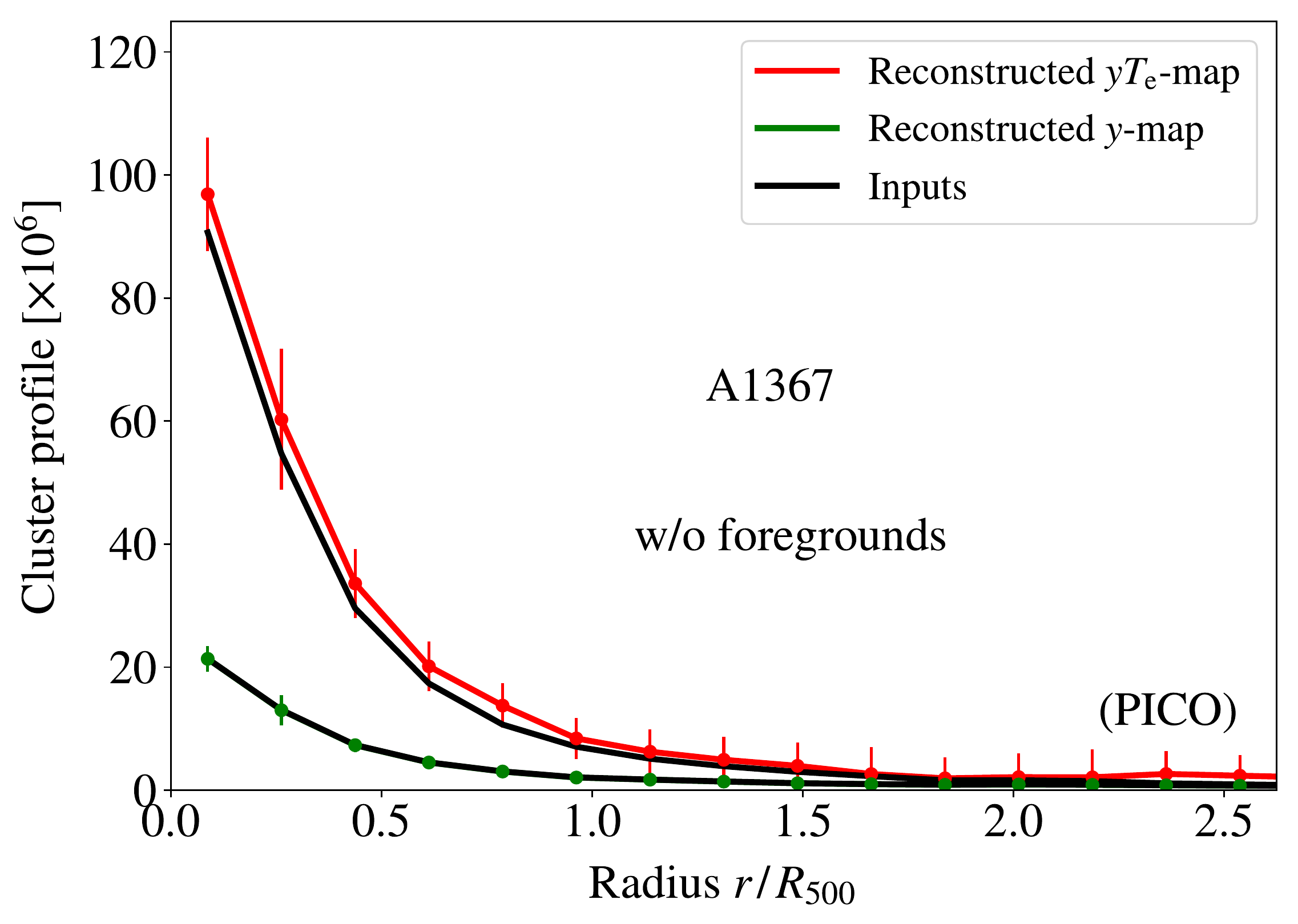}
    \\[0.5mm]
    \includegraphics[width=0.97\columnwidth]{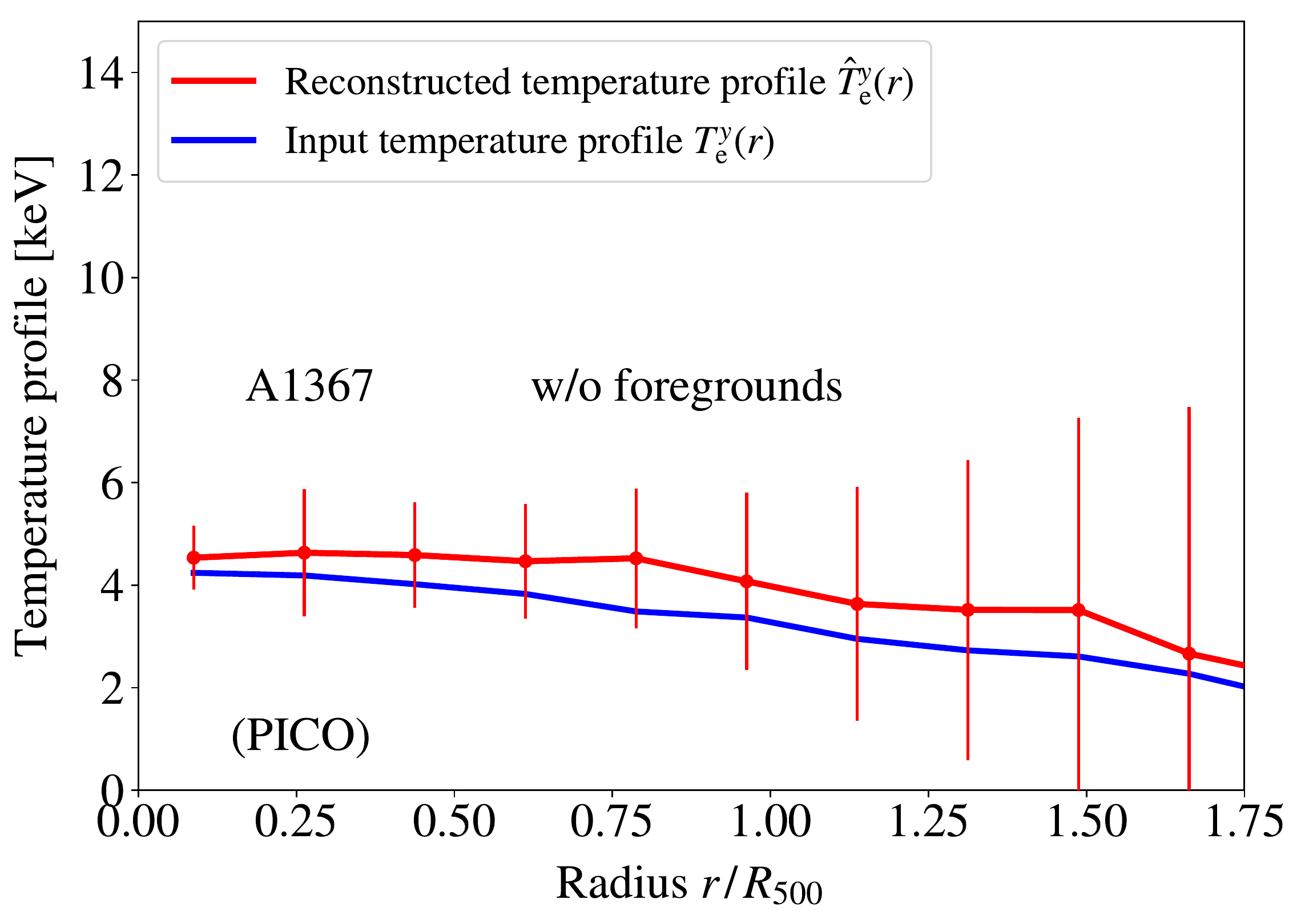}
  \end{center}
\caption{{ \pico\ without foregrounds.} \textit{Top}: Abell~1367 cluster profiles from the reconstructed $y$-map (\textit{green}) and the reconstructed $y\Te$-map (\textit{red}). \textit{Bottom}: Reconstructed temperature profile of Abell~1367 cluster from the ratio of the $y\Te$ profile and the $y$ profile (Eq.~\ref{eq:te2}). }
\label{Fig:profiles_and_power_nofg2}
\end{figure}

\begin{figure}
  \begin{center}
    \includegraphics[width=0.97\columnwidth]{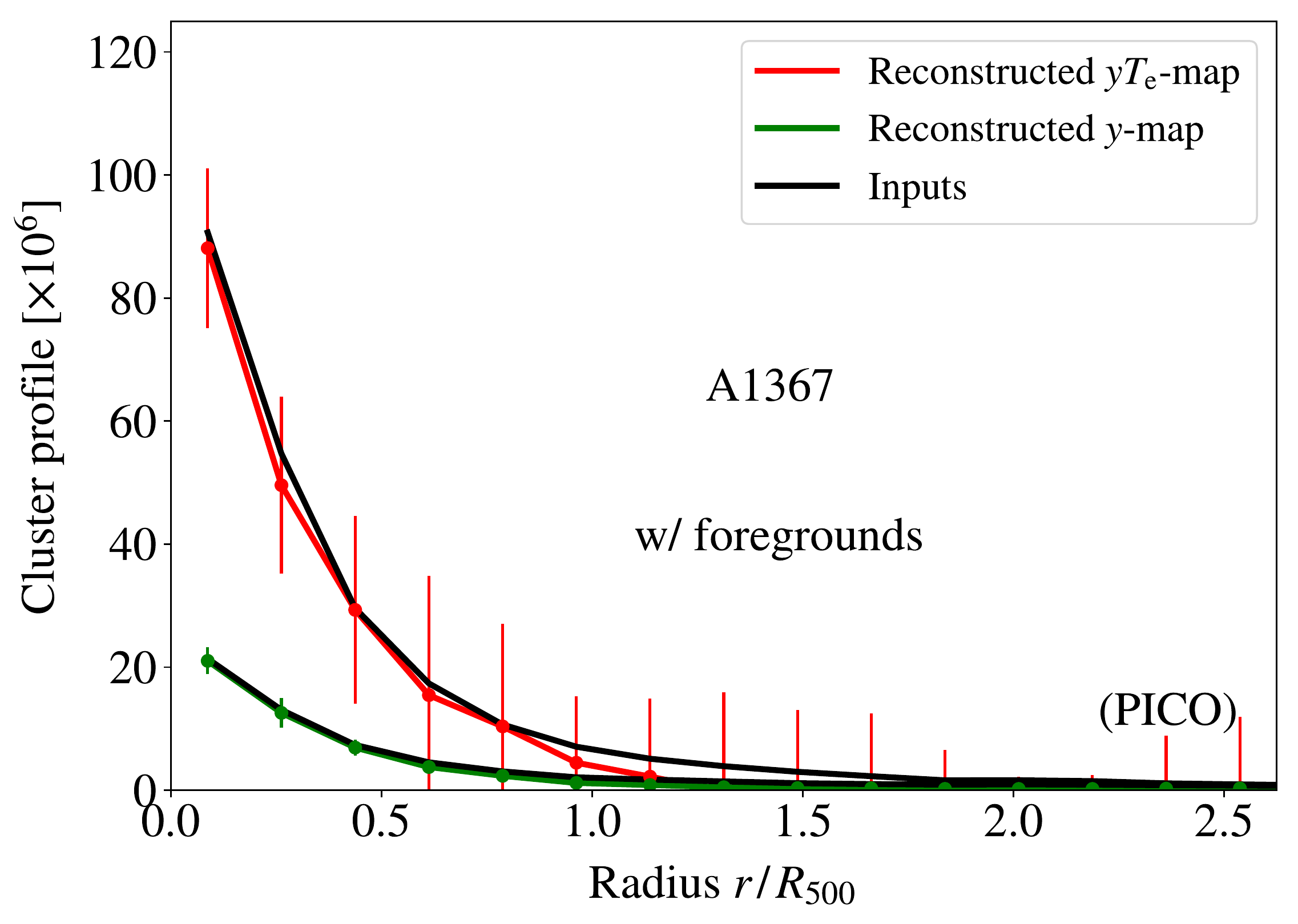}
    \\[0.5mm]
    \includegraphics[width=0.97\columnwidth]{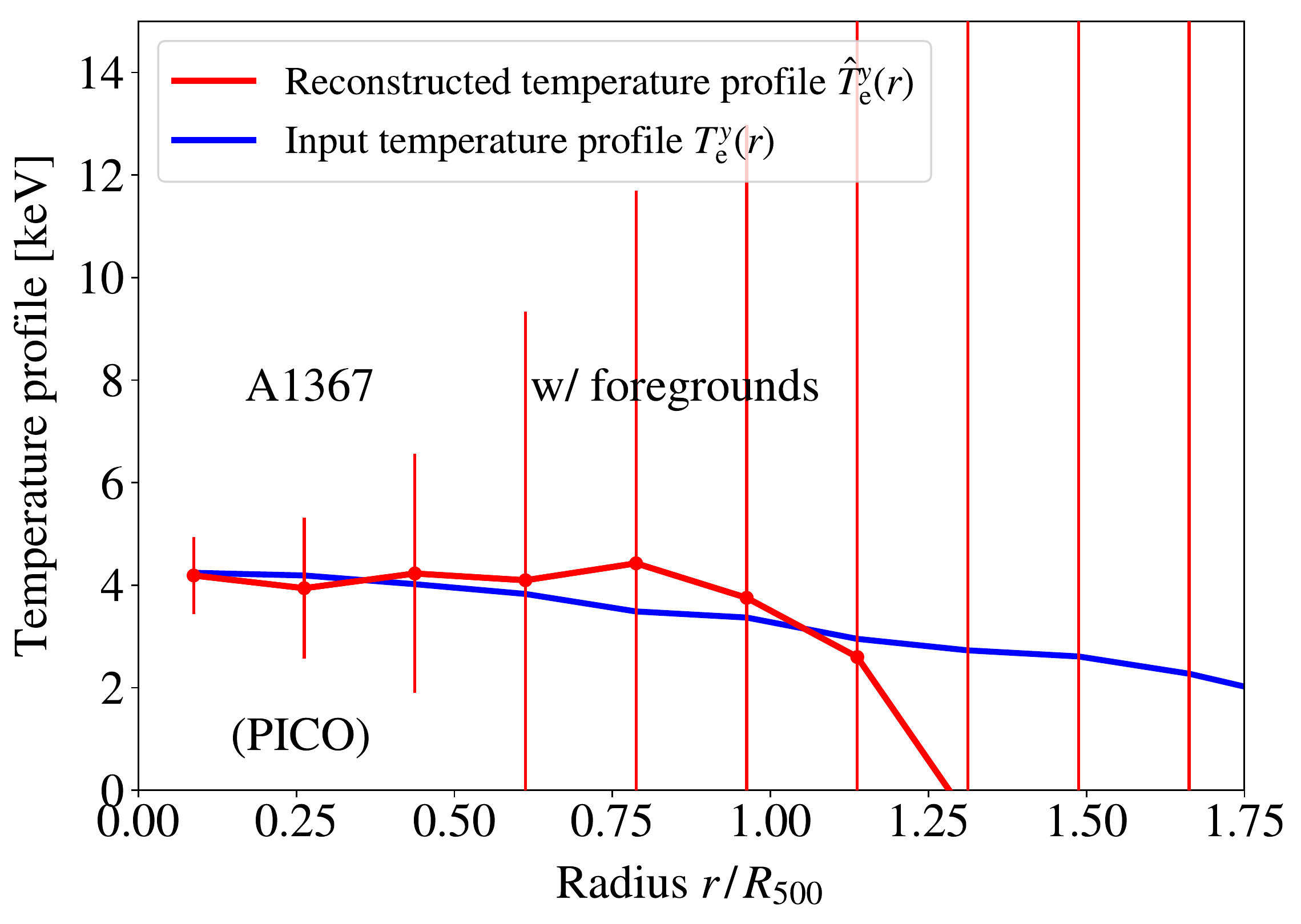}
  \end{center}
\caption{{ \pico\ with foregrounds.} \textit{Top}: Abell~1367 cluster profiles from the reconstructed $y$-map (\textit{green}) and the reconstructed $y\Te$-map (\textit{red}). \textit{Bottom}: Reconstructed temperature profile of Abell~1367 cluster from the ratio of the $y\Te$ profile and the $y$ profile (Eq.~\ref{eq:te2}).}
\label{Fig:profiles_and_power2}
\end{figure}

Aside from full-sky temperature mapping, we also show the reconstruction of thermodynamical profiles of individual clusters in Figs.~\ref{Fig:profiles_and_power_nofg}-\ref{Fig:profiles_and_power_nofg2} (\textit{without foregrounds}) and Figs.~\ref{Fig:profiles_and_power}-\ref{Fig:profiles_and_power2} (\textit{with foregrounds}). Profiles are computed as follows: we define non-overlapping successive rings of $1.5'$ pixels with increasing inner and outer radius, and we average the signal amplitudes, either $y$ or $y\Te$, over the pixels contained in a given ring. Error bars are obtained by computing the standard deviation of the signal fluctuations in each ring of pixels. The radius bins (or ring widths) are fixed to the beam FWHM value of the reconstructed maps ($5'$ for \pico) in order to minimise correlated errors between bins. In the top panels of Figs.~\ref{Fig:profiles_and_power_nofg}-\ref{Fig:profiles_and_power_nofg2} and Figs.~\ref{Fig:profiles_and_power}-\ref{Fig:profiles_and_power2}, we have plotted the cluster profiles of Coma and Abell~1367 for both the reconstructed $y$-map (green line) and the reconstructed $y\Te$-map (red line), with or without foregrounds. The quality of the profile reconstruction from \pico\ data is preserved in the presence of foregrounds (Fig.~\ref{Fig:profiles_and_power} versus Fig.~\ref{Fig:profiles_and_power_nofg}). The two reconstructed profiles $y(r)$ and $y\Te(r)$ match the input profiles (black lines) within error bars up to quite large radii from the centre of the cluster. 

In Figs.~\ref{Fig:profiles_and_power_nofg}-\ref{Fig:profiles_and_power_nofg2} (\textit{without foregrounds}) and Figs.~\ref{Fig:profiles_and_power}-\ref{Fig:profiles_and_power2} (\textit{with foregrounds}) we have computed the reconstructed temperature profiles $\Tewh^{\,y}(r)$ of Coma and Abell~1367 through the ratio (Eq.~\ref{eq:te2}) between the two aforementioned cluster profiles (this result has parallels with the map result shown in the bottom panels of Fig.~\ref{Fig:maps} and Fig.~\ref{Fig:maps2}). Error bars on the temperature profiles are computed from the variance of a ratio between two random variables, ${\Tewh^{\,y}(r) = \hat{z}(r) / \hat{y}(r)}$, as given by Eq.~\eqref{eq:stddev_ratio}. The temperature profile of $\Te(r)\simeq 7$\,keV of Coma is recovered without bias and with relatively high precision up to a radius of $R_{500}$ after foreground cleaning (bottom panel of Fig.~\ref{Fig:profiles_and_power}). A small systematic bias still persists because higher order temperature moments around the pivot temperature have been neglected in Eq.~\eqref{eq:moment}, but it is still within uncertainties. Adding constraints on higher order moments may improve the accuracy of the reconstruction but at the cost of increased error bars. Taking the inverse-variance weighted mean, the recovered mean temperature of Coma in the absence of foregrounds is of ${\Tewh^{\,y}=(6.9\pm 0.53)}$\,keV within $R_{500}$ radius, while in the presence of foregrounds the recovered temperature is of  ${\Tewh^{\,y}=(7.1\pm 0.69)}$\,keV within $R_{500}$ radius, and of ${\Tewh^{\,y}=(7.1\pm 0.68)}$\,keV within $2R_{500}$ radius, hence a $10\sigma$ measurement of the temperature of Coma ($7$\,keV) within $R_{500}$ with \pico\ after foreground removal. 
The addition of foregrounds only degrades the error of the recovered temperature value by $\simeq 30\%$ using our map-based extraction method.

For the smaller cluster Abell~1367, the fiducial temperature of $5$\,keV is recovered within $R_{500}$ radius in the absence of foregrounds, with a mean temperature of ${\Tewh^{\,y}=(4.5\pm 0.41)}$\,keV (bottom panel of Fig.~\ref{Fig:profiles_and_power_nofg2}). In the presence of foregrounds, the recovered mean temperature of Abell~1367 is of ${\Tewh^{\,y}=(4.1\pm 0.63)}$\,keV within $R_{500}$ radius after foreground removal (bottom panel of Fig.~\ref{Fig:profiles_and_power2}).  At large radii, the uncertainty on the recovered electron temperature profile of Abell~1367 increases due to residual foreground contamination. These important results show that future CMB experiments like \pico\ will provide the required sensitivity against foregrounds for our component separation method to measure the relativistic electron temperatures of several hundreds of galaxy clusters in the sky (Fig.~\ref{Fig:all-sky-te}), allowing furthermore to break the degeneracy between electron density and temperature in the cluster pressure profiles. In turn, the measurement of the electron density profiles should allow us to break the degeneracy in kinetic SZ measurements and access the cluster peculiar velocities without relying on external data sets.

\begin{figure}
  \begin{center}
    \includegraphics[width=0.97\columnwidth]{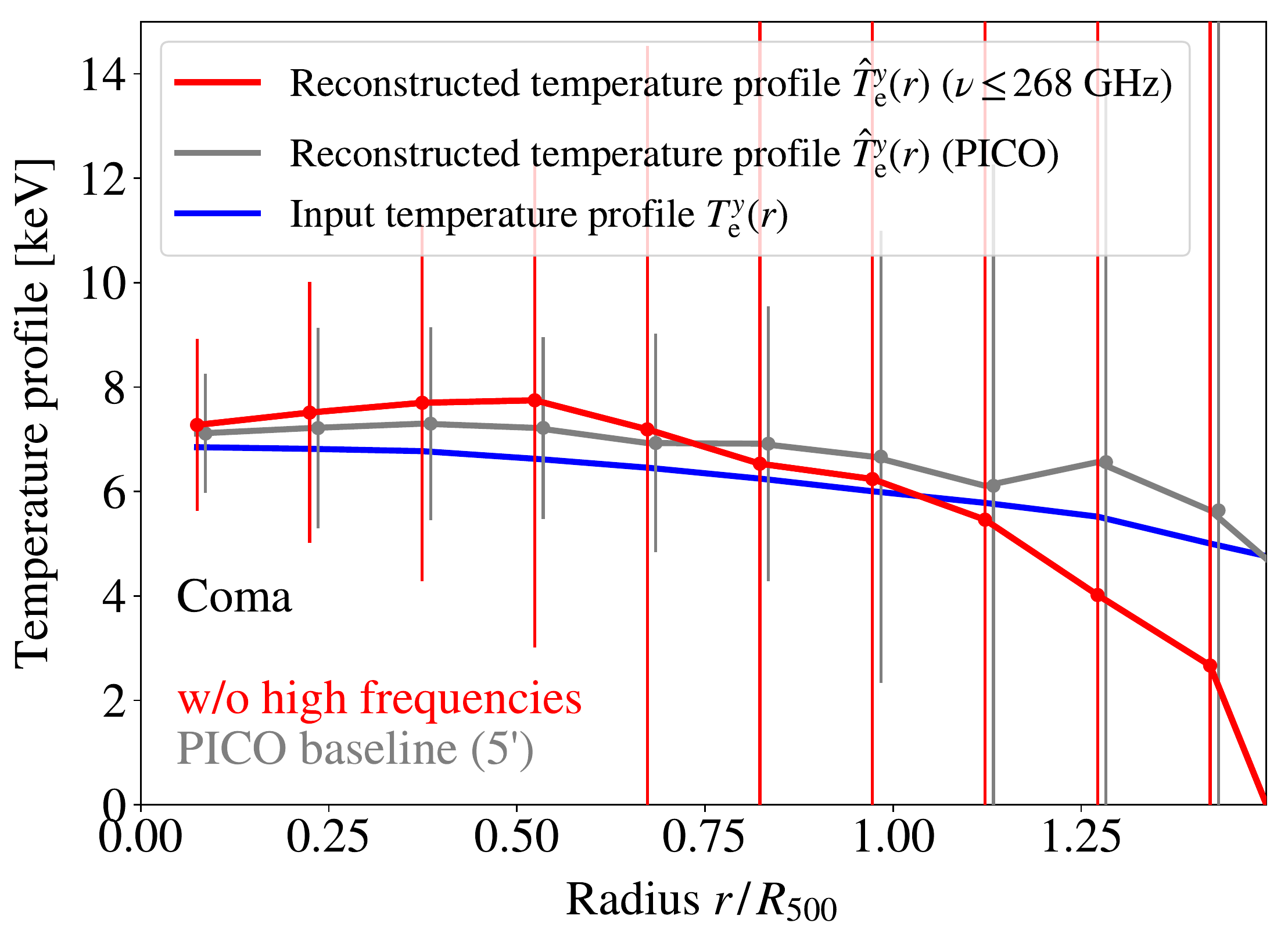}
  \end{center}
\caption{{ Reconstructed temperature profile of Coma with \pico\ ($5'$ resolution) when discarding high frequencies $\nu > 280$\,GHz (\textit{red line}) versus \pico\ baseline (\textit{grey line})}. }
\label{Fig:no_high_nu}
\end{figure}

Aside from helping in cleaning thermal dust foregrounds, high-frequency observations $> 300$\,GHz in future CMB experiments are essential to break temperature degeneracies in the relativistic thermal SZ spectrum \citep[e.g.,][]{Astro20202019Basu,voyage2050_backlight}, and hence to measure the actual temperature of clusters without confusion. 
For illustration, we discard all \pico\ frequencies above $280$\,GHz. The reconstruction of the temperature profile $\Te^y(r)$ of Coma in this case is shown in Fig.~\ref{Fig:no_high_nu} (red line), and should be compared to the baseline \pico\ configuration with full frequency coverage (grey line: copy from the bottom panel of Fig.~\ref{Fig:profiles_and_power}). Clearly, in the absence of channels at frequencies $> 280$\,GHz, the recovered temperature of Coma has much larger uncertainties and cannot be distinguished from a zero temperature at radii $r > 0.5R_{500}$, because of temperature degeneracies in the relativistic thermal SZ spectrum at low frequencies. 
Without high frequencies, the recovered mean temperature of Coma is degraded to $\Tewh^y=(7.4\pm 1.2)$\,keV within $R_{500}$, while for full frequency coverage of the baseline \pico\ configuration the recovered temperature was of $\Tewh^y=(7.1\pm 0.69)$\,keV. This corresponds to a degradation by more than $70$\,\% although channels at $\nu > 280$\,GHz only make up $30\%$ of the \pico\ frequency bands. For galaxy clusters fainter and cooler than Coma, we could expect no detection of the cluster temperature at all for CMB experiments that do not probe high frequencies $\nu > 280$\,GHz. 
The experimental requirement of high frequencies for measuring relativistic SZ temperatures should be put in contrast to the case of $\mu$-type spectral distortions \citep[e.g.,][]{Sunyaev1970mu, Burigana1991, Hu1993, Chluba2011therm}, for which low-frequency channels were found to give more constraining power than high frequencies due to the specific spectral shape of the $\mu$-distortion \citep{abitbol_pixie,Remazeilles2018MNRAS}. A combination of ground-based and space-based experiments may thus provide a promising avenue for overcoming these limitations, allowing us to optimize sensitivity, angular resolution and frequency-coverage.

\subsubsection{$\Te$ power spectrum}\label{subsubsec:Te-spectrum}

\begin{figure}
  \begin{center}
       \includegraphics[width=0.97\columnwidth]{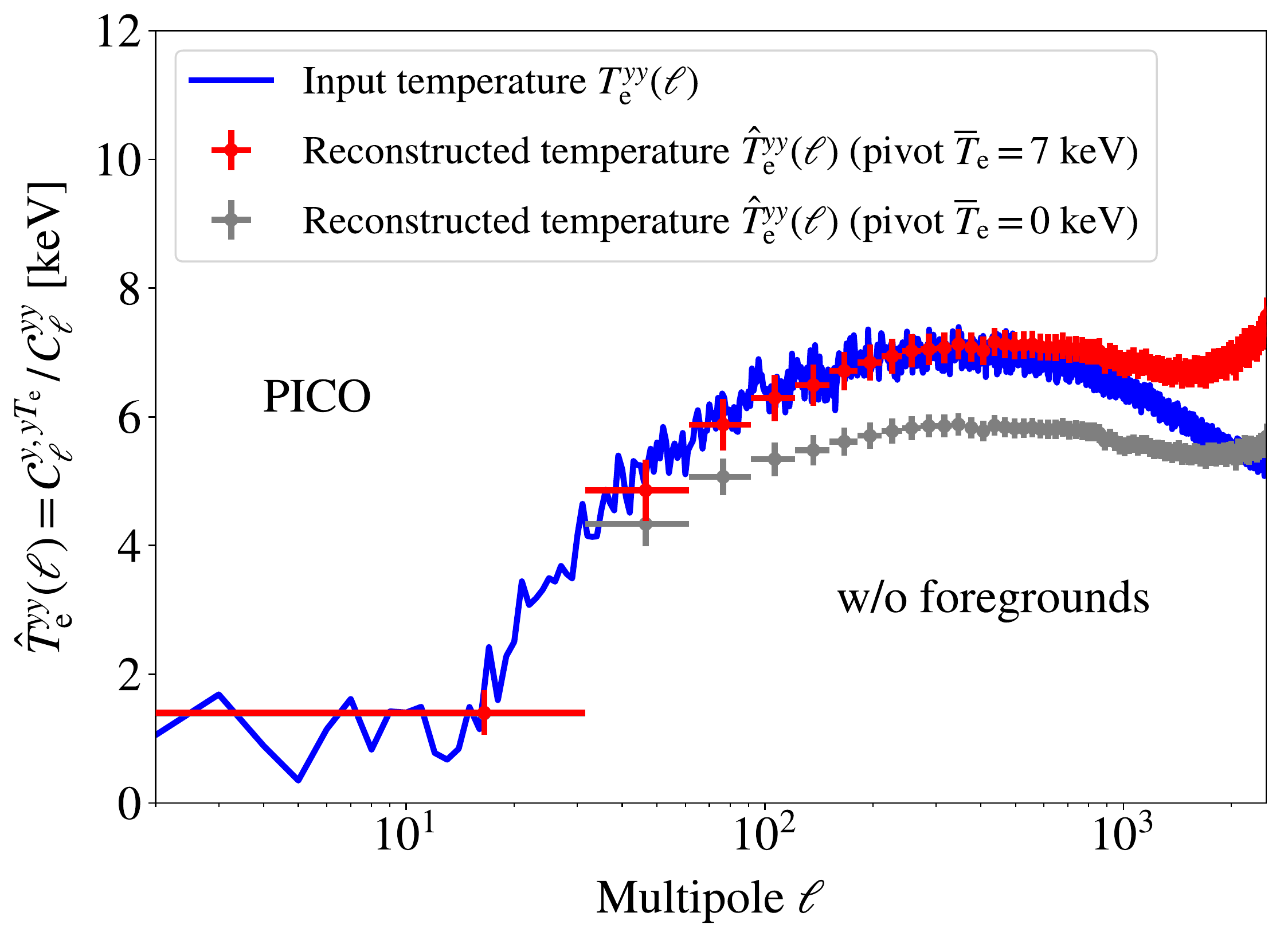}
       \\[2mm]
  	 \includegraphics[width=0.97\columnwidth]{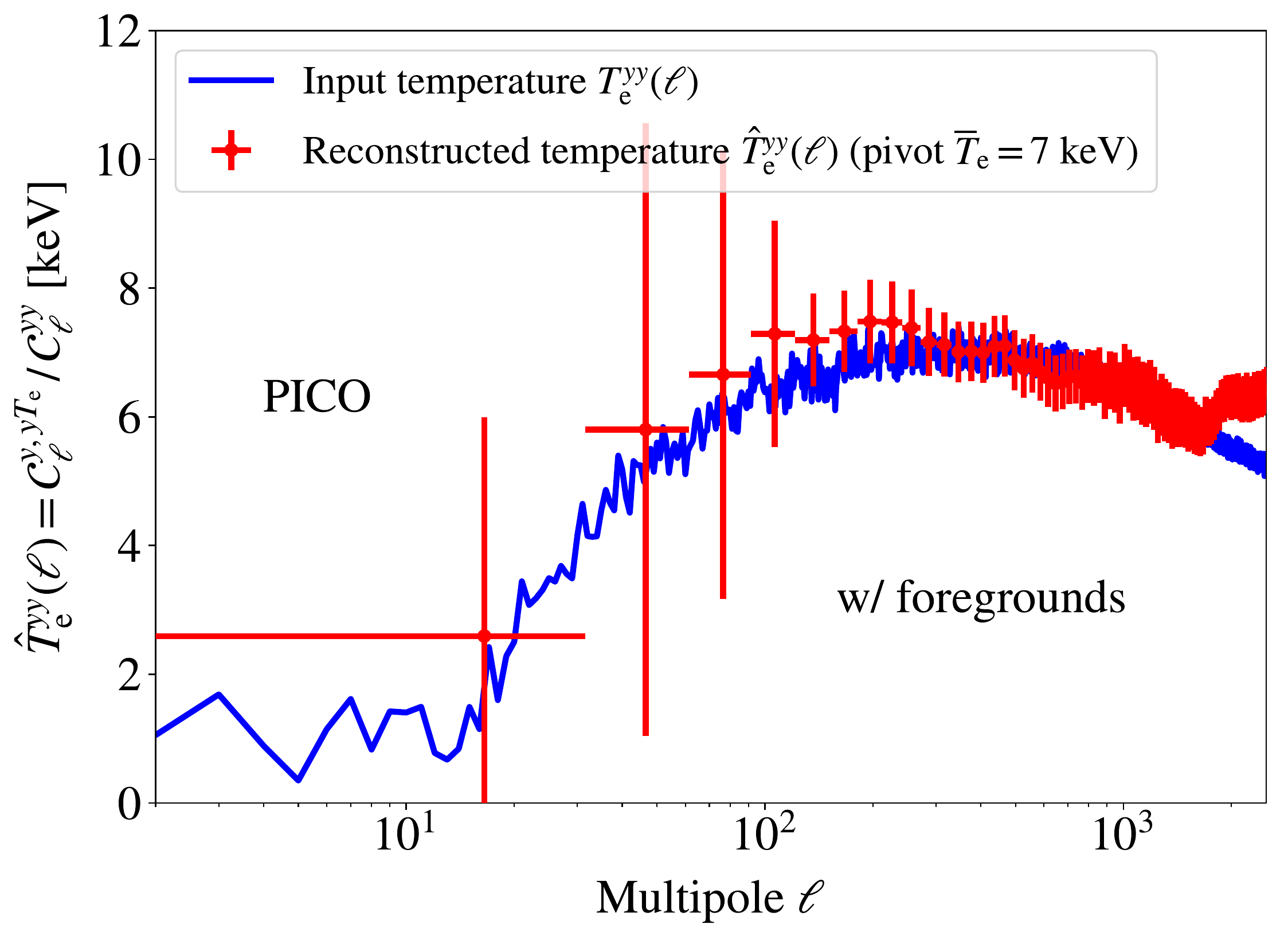}
  \end{center}
\caption{{Reconstructed ${y^2}$-weighted average temperature (Eq.~\ref{eq:te1}) over the sky for \pico}. A uniform pivot temperature of $\Teol=7$\,keV over angular scales has been adopted in the analysis. \textit{Top}: without foregrounds. \textit{Bottom}: with foregrounds.}
\label{Fig:average_te}
\end{figure}

Finally, we investigate the reconstruction of the \textit{average} electron temperature over the sky across several angular scales (Fig.~\ref{Fig:average_te}). Cosmological studies of galaxy clusters for cosmological parameter inference rely on the properties of a significant ensemble of clusters through either cluster number counts \citep{Planck2015clustercount} or SZ power spectrum analysis \citep{Planck2015ymap}, for which relativistic corrections to SZ effects will become more significant \citep{Erler2018,Remazeilles2019}. 
The definition of the \textit{average}  temperature of clusters that is the most relevant depends on the specific cosmological observable (cluster number counts, stacking, power spectrum, or monopole), and several definitions have been used in the literature: mass-weighted average temperature, $\tau$-weighted average temperature (where $\tau$ is the optical depth), or $y$-weighted average temperature. As shown in \cite{Remazeilles2019}, the relevant average cluster temperature for SZ power spectrum analysis is the scale-dependent $y^2$-weighted average temperature (see Eq.~\eqref{eq:te1} in the present paper) because it cancels the linear bias in $\Te$ in the thermal SZ power spectrum.

In Fig.~\ref{Fig:average_te}, we computed the cross-power spectrum between the two reconstructed $y$- and $y\Te$-maps using pseudo-$C_\ell$ estimators \citep{Wandelt2001}, and divided it by the auto-power spectrum of the reconstructed $y$-map in order to estimate the $y^2$-weighted average temperature $\Tewh^{\,yy}(\ell)$ (Eq.~\ref{eq:te1}) after component separation. The power spectra $C_\ell^{y,y\Te}$ and $C_\ell^{yy}$ are binned through regular bins of $\Delta\ell = 30$, while the error bars are computed analytically \citep[e.g.][]{Tristram2005}:
\begin{subequations}
\begin{align}
\label{eq:error-spectra}
&\sigma\left( C_{\ell_b}^{yy} \right) = \sqrt{ \frac{2\,\left(C_{\ell_b}^{yy}\right)^2 }{ \left(2\ell_b+1\right)\, \Delta\ell\, f_{\rm sky}} },
\\
\label{eq:error-spectra2}
&\sigma\left( C_{\ell_b}^{y,y\Te} \right) = \sqrt{ \frac{C_{\ell_b}^{yy}\,C_{\ell_b}^{y\Te,y\Te}\,+\,\left(C_{\ell_b}^{y,y\Te}\right)^2 }{ \left(2\ell_b+1\right)\, \Delta\ell\, f_{\rm sky}} },
\\
\label{eq:error-spectra3}
&{\rm Cov}\left(C_{\ell_b}^{yy}, C_{\ell_b}^{y,y\Te} \right) = \frac{2}{ \left(2\ell_b+1\right)\, \Delta\ell\, f_{\rm sky}}\,C_{\ell_b}^{yy}\,C_{\ell_b}^{y,y\Te},
\end{align}
\end{subequations}
where $\ell_b$ is the central multipole of the bin and $f_{\rm sky}$ is the fraction of the sky outside the Galactic mask. Equations~\eqref{eq:error-spectra}-\eqref{eq:error-spectra2}-\eqref{eq:error-spectra3} thus includes the cosmic variance of the signal in addition to the variance from residual foregrounds in the reconstructed power spectra after component separation. The resulting error bar of $\Tewh^{\,yy}(\ell_b) = C_{\ell_b}^{y,y\Te}\,/\,C_{\ell_b}^{yy}$ is then obtained using Eq.~\eqref{eq:stddev_ratio}, which does not suffer much from cosmic variance limitation since $\Tewh^{\,yy}$ is the ratio of the cross- and auto-power spectra of two biased tracers, $y$ and $y\Te$, that come from the same realisation of the underlying matter field \citep{Seljak2009,Witzemann2019}.

 In the absence of foregrounds (top panel of Fig.~\ref{Fig:average_te}), the reconstructed average temperature $\Tewh^{\,yy}(\ell)$ (red line) for a uniform pivot temperature of $\Teol=7$\,keV across multipoles is consistent with the input average temperature of the cluster simulation (blue line) over a large range of multipoles $10\lesssim \ell \lesssim 1000$. The increase at $\ell > 1000$ is due to residual noise. Interestingly, the shape of $\Tewh^{\,yy}(\ell)$ in our simulations (blue line), which are based on public ROSAT and SDSS cluster catalogues, is consistent with our earlier theoretical expectations \citep[see Fig.~3 in][]{Remazeilles2019}. We note that had we adopted a pivot temperature of $\Te=0$ in our analysis then the recovered electron temperature would be underestimated across the same broad range of angular scales (grey line), hence showing the importance of our temperature spectroscopy and moment expansion around non-zero pivot temperatures. 
 
In the presence of foregrounds (bottom panel of Fig.~\ref{Fig:average_te}), we are still able to recover the overall behaviour of the fiducial average SZ temperature power spectrum over a broad range of angular scales, $100\lesssim \ell \lesssim 1500$. However, at low multipoles $\ell < 100$ (or large angular scales) the recovery degrades due to residual foregrounds. 
The increase of uncertainties on $\Te^{\,yy}(\ell)$ at large angular scales is threefold: first, the averaging effect of different line-of-sight electron temperatures is more significant at large angular scales, in a way that will change the effective spectral response of the relativistic thermal SZ effect \citep{Chluba2012moments, Chluba2017}, therefore increasing the uncertainty on the recovered temperature. 
Second, residual foreground contamination is larger when averaged over large angular scales, therefore further increasing the uncertainties on the recovered temperatures at low multipoles.
Both aspects could be improved by adding higher order moment constraints to the SZ and dust signals.
Third, the reconstruction of the scale-dependent average SZ temperature in Fig.~\ref{Fig:average_te} was performed by assuming a uniform pivot temperature of $\Teol=7$\,keV over the whole range of angular scales in our analysis. This is a good approximation given that the average value of $\Te^{yy}(\ell)$ over the angular scales $100\lesssim \ell \lesssim 1000$  in our simulations (blue line in Fig.~\ref{Fig:average_te}) is of $\langle \Te^{yy}(\ell) \rangle_\ell \simeq 7$\,keV. 
However, the reconstruction could be optimised at large angular scales by adopting varying pivot temperatures $\Teol\equiv \Teol(\ell)$ across multipole ranges (or needlet scales) in the component separation analysis. This modification is expected to reduces residual foreground projections currently found by our method at the largest scales.
A possible prior on the pivot values $\Teol(\ell)$ to be used could be taken from the theoretical estimate $\Te^{yy}(\ell)$ derived in \cite{Remazeilles2019}. Since our analysis is performed on a needlet frame, we could integrate the expected theoretical $\Te^{yy}(\ell)$ over each needlet window (bandpass windows across multipoles in Fig.~\ref{Fig:needlets}) to get the appropriate pivot temperatures over each relevant range of angular scales. For this purpose, the needlet windows will have to be chosen carefully in order to minimise the gradients of temperature in each range of multipoles. This will be investigated in the future. 

To conclude, we stress the following important points. Our component separation method delivers a new map-based observable, $\Te$, for cosmology, in addition to the usual Compton-$y$ parameter observable. The shape and amplitude of the temperature power spectrum $\Te^{\,yy}(\ell)$ depends on the underlying cosmological parameters and cluster physics, but in a different way than the Compton-$y$ power spectrum observable, $C_\ell^{\,yy}$. In particular, $\Te^{\,yy}(\ell)$ is expected to vary with $\sigma_8$ but should be mostly insensitive to the mass-bias $(1-b)$ by construction \citep{Remazeilles2019}, unlike $C_\ell^{\,yy}$ and cluster number counts. 
To provide some intuition on our expectation, let us consider the theoretical form of the Compton-$y$ power spectrum \citep{Komatsu:2002wc}:
\begin{align}
\label{eq:theory}
&C_{\ell}^{yy}  = \int_{\,0}^{\,z_{\rm max}} dz{dV\over dz}  \int_{\,M_{\rm min}}^{\,M_{\rm max}} dM\, {dn(M,z) \over dM}\,\vert y_\ell (M,z)\vert^2,
\end{align}
where ${dn(M,z) / dM}$ is the dark matter halo mass function, and $y_\ell (M,z)$ the two-dimensional Fourier transform of the projected Compton-$y$ radial profile. Similarly, we have:
\begin{align}
\label{eq:theory2}
&C_{\ell}^{y,y\Te}  = \int_{\,0}^{\,z_{\rm max}} dz{dV\over dz}  \int_{\,M_{\rm min}}^{\,M_{\rm max}} dM\, {dn(M,z) \over dM}\,\Te(M,z)\,\vert y_\ell (M,z)\vert^2,
\end{align}
which has the same dependence as $C_{\ell}^{yy}$ on the projected $y$-profile through the term $\vert y_\ell (M,z)\vert^2$, but a different effective halo mass function through the modulation by the temperature $\Te(M,z)$. Therefore, $C_{\ell}^{yy}$ and $C_{\ell}^{y,y\Te}$ have the same scaling dependence with the mass bias, which enters in the $y$-profile term $\vert y_\ell (M,z)\vert^2$ only through the Y-M scaling relation, while they have a different scaling with $\sigma_8$ since the halo mass function in $C_{\ell}^{y,y\Te}$ is modulated by the temperature $\Te(M,z)$ which depends on $\sigma_8$ through the temperature-mass (T-M) scaling relation. Being defined as the ratio of $C_{\ell}^{y,y\Te}$ and $C_{\ell}^{yy}$, the relativistic temperature power spectrum $\Te^{\,yy}(\ell)$ (Eq.~\ref{eq:te1}) will thus still depend on $\sigma_8$, while being mostly insensitive to the mass bias, since the main scaling with the mass bias is broken by the ratio. Therefore, such a new SZ map-based observable, $\Te^{\,yy}(\ell)$, offers an additional degree of freedom to constrain cosmological parameters and break degeneracies in future SZ cosmological analyses, something we plan to explore more in depth.

\vspace{0mm}
\subsection{Large-scale diffuse electron temperature and Coma's temperature with \litebird}\label{subsec:litebird}
The telescope mirror of the CMB space mission \litebird\ \citep{Suzuki2018} is significantly smaller than the telescope of \pico, with an average optical beam resolution of \litebird\ over the frequency bands of $\simeq 30$\, arcmin. While the primary focus of \litebird\ is not on cluster science, the large sensitivity of this experiment and its relatively broad frequency coverage ($40$--$402$\,GHz) may still allow us to measure the large-scale diffuse electron temperature in the sky, as well as the temperature of the largest clusters like Coma. 

\begin{figure}
  \begin{center}
       \includegraphics[width=0.97\columnwidth]{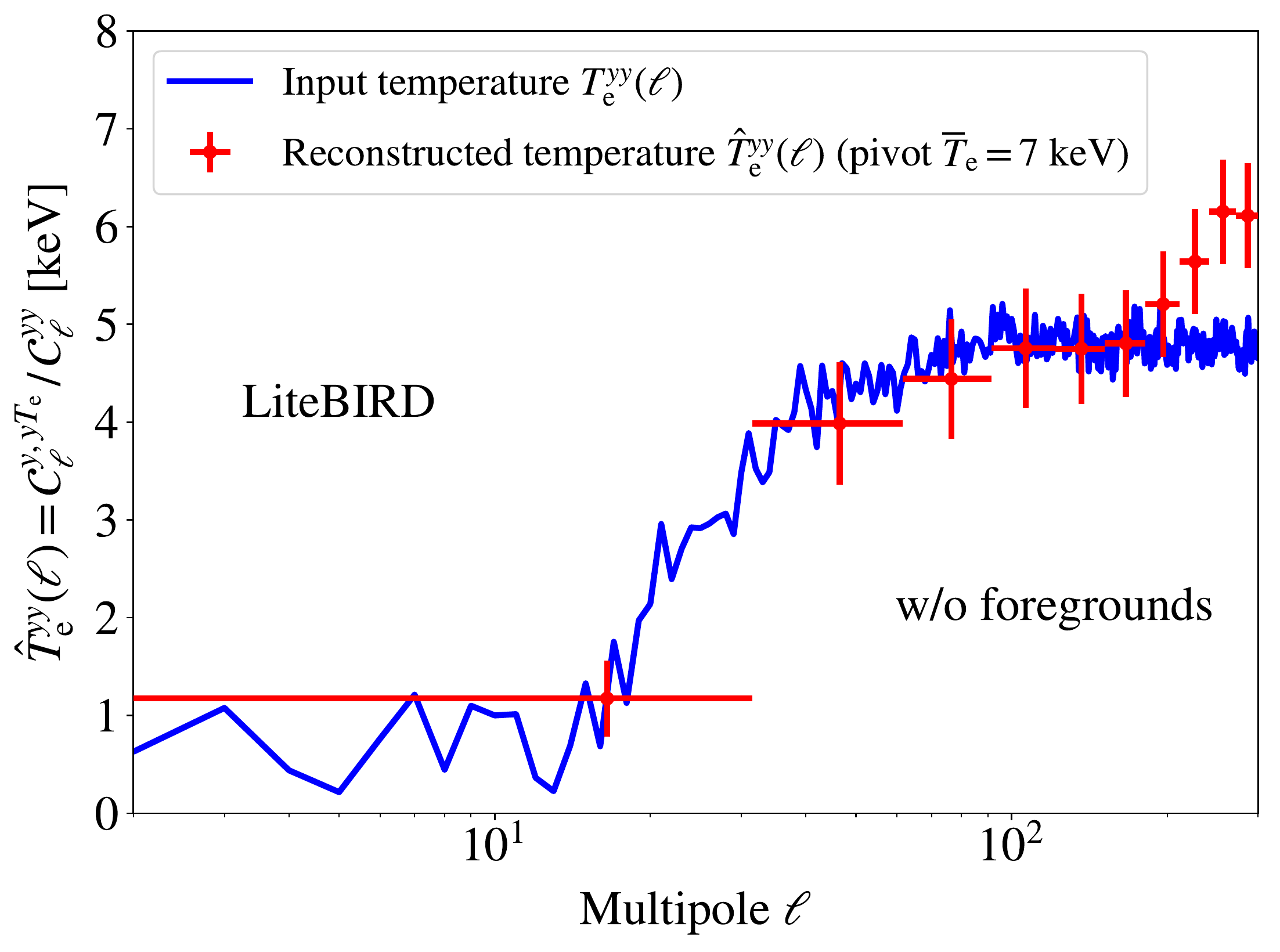}
  \end{center}
\caption{Reconstructed ${y^2}$-weighted average temperature (Eq.~\ref{eq:te1}) over the sky for \litebird, \textit{in the absence of foregrounds}. A uniform pivot temperature of $\Teol=7$\,keV over angular scales has been adopted in the analysis.}
\label{Fig:average_te_ltb}
\end{figure}

We thus explored the performance of \litebird\ experiment in recovering the $y^2$-weighted average temperature over the sky. As shown in Fig.~\ref{Fig:average_te_ltb}, in the absence of foregrounds \litebird\ would manage to recover the average electron temperature over the sky at large angular scales $10\lesssim \ell \lesssim 200$, while temperatures at smaller scales cannot be measured due to lack of resolution. The overall temperature (blue line) for the \litebird\ simulation is lower compared to the \pico\ simulation because most clusters are unresolved and the temperatures across the sky are diluted by the larger pixelization of low-resolution \litebird\ observations.

In the presence of foregrounds, we found that the recovered temperatures for \litebird\ are consistent with zero across the same range of angular scales. This degradation can be due to the combination of lack of high frequencies and resolution, as stressed earlier in Fig.~\ref{Fig:no_high_nu}. Adding a few frequencies above $400$\,GHz to \litebird\ could enhance its performance in measuring the large-scale diffuse electron temperature in the presence of foregrounds. However, there may also be some room for improvement of the component separation method to reduce residual foreground contamination down to the noise levels of Fig.~\ref{Fig:average_te_ltb}. Adding a constraint to Eqs~\eqref{eq:constbis-1}-\eqref{eq:constbis-4} against, e.g., the first moment of the thermal dust spectrum, ${\partial\, [ \nu^{\,\beta_d}\,B(\nu,T_d) ]\,  /\, \partial\, \beta_d}\, \vert_{\,\beta_d=\overline{\beta}_d}$, could help in removing dust residuals. This has to be explored carefully since adding constraints is at the cost of increasing the residual noise in the reconstruction, hence we have to find the optimal point of minimum residuals between fully blind and fully parametric approaches, something we plan to investigate in the future.

\begin{figure}
  \begin{center}
    \includegraphics[width=0.97\columnwidth]{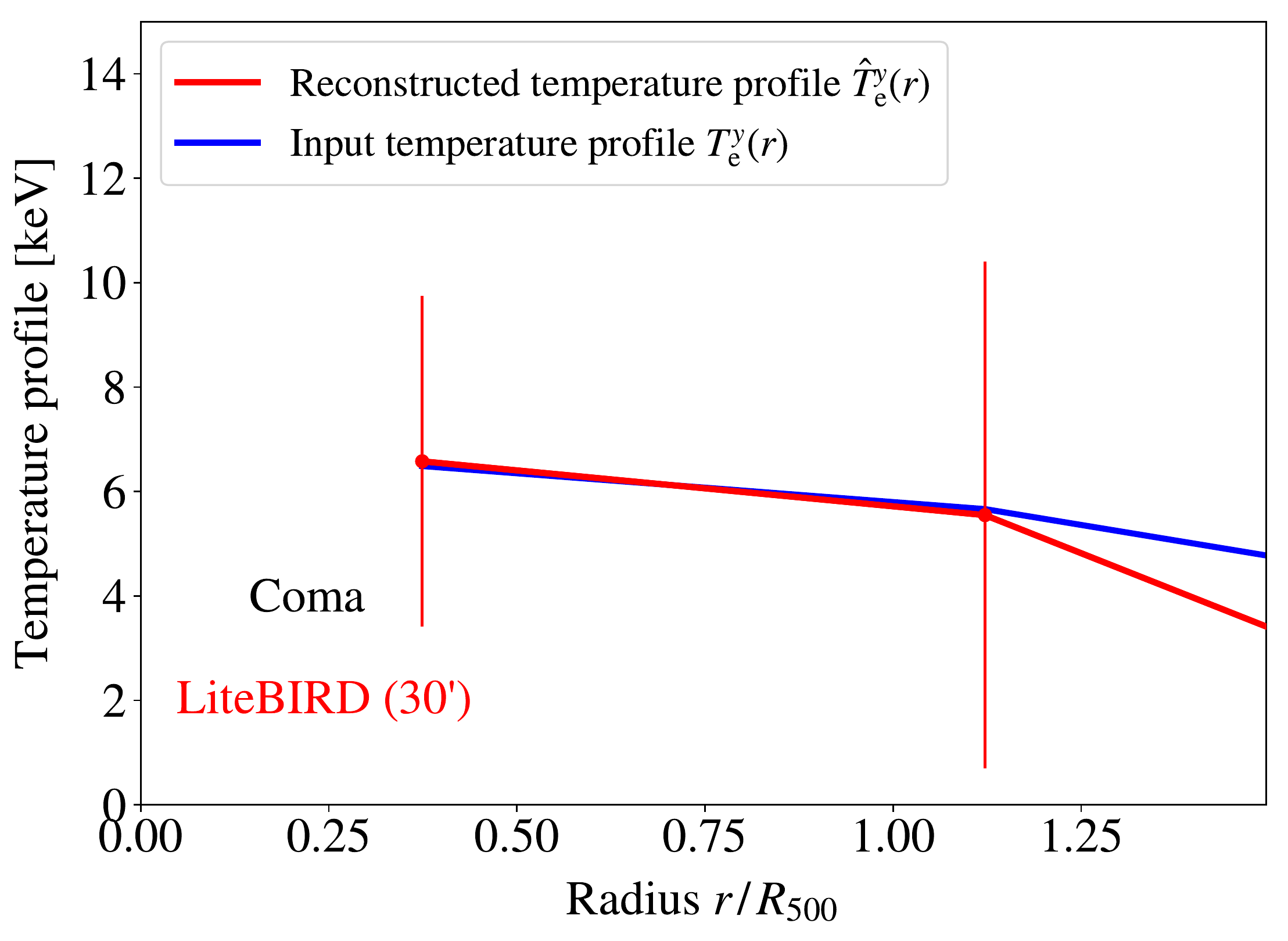}
  \end{center}
\caption{Reconstructed temperature profile of Coma with \litebird\ ($30'$ resolution), \textit{in the presence of foregrounds}.}
\label{Fig:litebird}
\end{figure}

In Fig.~\ref{Fig:litebird}, we show the reconstructed temperature profile of Coma at 30' resolution for \litebird\ in the presence of foregrounds. The low spatial resolution of \litebird\ results in large bins of temperature. This result should be compared with the result for \pico\ shown in the bottom panel of Fig.~\ref{Fig:profiles_and_power} or with the grey line of Fig.~\ref{Fig:no_high_nu}. While the uncertainty on the temperature measurement is much larger for \litebird\ and the resolution lower, we can still measure the average temperature of Coma after foreground removal, which we find to be 
${\Tewh^{\,y} = (6.0 \pm 2.6)}$\,keV within $2R_{500}$, hence about a $2\sigma$-$3\sigma$ measurement, where the same result for \pico\ was ${\Tewh^{\,y} = (7.1 \pm 0.68)}$\,keV, i.e. a $10\sigma$ measurement. To conclude, even if not optimized for SZ science, \litebird\ is still a very promising experiment for probing the temperature power spectrum of the diffuse electron gas at large angular scales and the temperature of the largest clusters in the sky.

\section{Conclusions}
\label{sec:conc}

We developed a new approach based on SZ-temperature moment expansion and Constrained-ILC component separation that allows mapping relativistic electron temperatures of galaxy clusters across the entire sky. This delivers a new map-based observable, $\Te$, in addition to the Compton-$y$ parameter for cosmological and astrophysical studies of galaxy clusters. The novel feature of our component separation method is that it also offers a spectroscopic view of the galaxy clusters not only across frequency but also across temperature (Sect.~\ref{subsec:spectroscopy}). We have shown that a next-generation CMB experiment like \pico\ provides the required sensitivity, angular resolution, and frequency coverage to map and measure the electron gas temperature of galaxy clusters across the entire sky, as well as constrain the average electron gas temperature over a broad range of angular scales, delivering the $T^{\,yy}_e(\ell)$ spectrum (see Fig.~\ref{Fig:average_te}). The shape and amplitude of this new map-based observable, $T^{\,yy}_e(\ell)$, depends on the underlying cosmology and cluster physics, as already stressed in \cite{Remazeilles2019}. 
We expect that this new observable will offer a new way to constrain cosmological parameters, cluster physics and break parameter degeneracies in the future.

In addition, our method allows us to blindly reconstruct the pressure and temperature profiles of individual clusters with high significance. 
This is particularly useful to future SZ data analysis since it may allow us to break the degeneracy between electron density and temperature in the cluster pressure profile. As a consequence, the obtained electron density profiles will help us to directly measure the peculiar velocities of individual galaxy clusters using a combination with the kSZ effect. 

We find that for Coma, \pico\ will measure its average electron temperature to $\simeq 10\sigma$ significance after foreground removal. Low-angular resolution CMB experiments like \litebird\ could still achieve $2\sigma$ to $3\sigma$ measurement of the electron temperature of this largest cluster (see Sect.~\ref{subsec:litebird}). Concerning the diffuse electron gas temperature, $\Te^{\,yy}(\ell)$, \litebird\ shows encouraging results at large angular scales based on noise-only simulations (Fig.~\ref{Fig:average_te_ltb}). However, it may require some further development on component separation or increased spectral coverage to achieve similar results in the presence of foregrounds.

For \pico, we also demonstrated that our map-based method can be used to reliably derive the temperatures of large samples of clusters across the sky (Fig.~\ref{Fig:all-sky-te}). This will provide a new way for determining cluster masses using the relativistic SZ effect. We refined the modelling of relativistic SZ effects by adopting moment expansions around non-zero pivot temperature instead of directly using the expressions of \cite{Itoh1998}, allowing for cluster spectroscopy across temperatures, and guaranteeing more reliable recovery of the actual temperatures of hot galaxy clusters.

We stress that our approach is fully \textit{map-based}. Therefore it differs from cluster extraction tools like matched multifilters \citep[MMF:][]{Herranz2002,Melin2006} in many interesting aspects. First, unlike MMF tools, our map-based method is blind to any cluster profile assumption, i.e., it does not rely on any universal form for the cluster pressure profile but it allows instead to blindly reconstruct individual cluster profiles. Second, our map-based method allows the reconstruction of the \textit{full} Compton-$y$ and temperature-$\Te$ fields, i.e., the electron temperature at cluster scales but also the large-scale diffuse electron gas temperature.

Our method provides also the first exploration of \textit{semi-blind} component separation analysis, in the sense that we set constraints against the statistical moments of some foregrounds, in contrast to astrophysical model fitting of fully parametric methods, hence drawing the path towards optimisation of component separation at the intersection of fully blind and fully parametric methods. While perfect knowledge of the spectrum of the component of interest is required to avoid biases (Eq.~\ref{eq:constbis-1}), we stress that partial knowledge of the spectrum of unwanted foregrounds through their moments is sufficient to remove bulk of the contamination (Eq.~\ref{eq:constbis-4}). By setting constraints against the zeroth moment of dust, the Constrained-ILC better removes dust than a simple variance minimisation of blind ILC methods, while avoiding biases of fully parametric methods that may arise from fitting incorrect astrophysical models \citep[as discussed in][]{Remazeilles2016}. The Constrained-moment-ILC method can in principle support additional constraints against higher-order moments of foregrounds, which we plan to explore further in the future to find out the optimal compromise between fully parametric and fully blind methods, where the gain in foreground removal by extra moment constraints overcomes the increase of variance.

Finally, we have shown how high-frequency observations and angular resolution are important for measuring relativistic cluster temperatures across the sky. In this respect, high-frequency surveys like \mbox{CCAT-prime} \citep{CCAT-p} and the space mission \mbox{Millimetron} \citep{Millimetron}, aside from \litebird\ and \pico, would be very useful for complementing the exquisite sensitivity and resolution of the Simons Observatory \citep{SO2019} and the CMB-S4 experiment \citep{CMB-S4_2016} at low frequency. Similarly, future CMB spectrometers like {\it PIXIE} \citep{Kogut2011PIXIE, Kogut2016SPIE}, with many hundreds of absolutely calibrated frequency channels, may be able to help in achieving many of these future goals.
It is worth noting that accurate channel inter-calibration will be essential for unbiased relativistic temperature measurements through component separation, while narrow bandpasses at high frequencies will be required to break temperature degeneracies in the relativistic thermal SZ spectrum. These systematics need to be carefully considered in future forecasts of relativistic SZ temperature spectroscopy.

\small

\vspace{-5mm}
\section*{Acknowledgements}
This project has received funding from the European Research Council (ERC) under the European Union's Horizon 2020 research and innovation programme (grant agreement No 725456, CMBSPEC). JC was also supported by the Royal Society as a Royal Society University Research Fellow at the University of Manchester. We thank \mbox{Keith Grainge}, \mbox{Shaul Hanany}, \mbox{Colin Hill}, \mbox{Eiichiro Komatsu}, and \mbox{Aditya Rotti} for useful suggestions and discussions about the project. We also thank Jacques Delabrouille and Jean-Baptiste Melin for useful discussions on SZ modelling in the PSM, and the anonymous referee for their comments and suggestions. Some of the results in this paper have been derived using the \healpix\ package \citep{Gorski2005}. We also acknowledge the use of the PSM package \citep{Delabrouille2013}, developed by the \Planck\ working group on component separation, for making the simulations used in this work.

\vspace{-5mm}
\bibliographystyle{mn2e}
\bibliography{rsz_temperature}

\bsp	
\label{lastpage}
\end{document}

%% file: Planck.tex
\def\setsymbol#1#2{\expandafter\def\csname #1\endcsname{#2}}
\def\getsymbol#1{\csname #1\endcsname}

\def\Planck{\textit{Planck}}

\newbox\tablebox    \newdimen\tablewidth
\def\leaderfil{\leaders\hbox to 5pt{\hss.\hss}\hfil}
\def\tablenote#1 #2\par{\begingroup \parindent=0.8em
    \abovedisplayshortskip=0pt\belowdisplayshortskip=0pt
    \noindent
    $$\hss\vbox{\hsize\tablewidth \hangindent=\parindent \hangafter=1 \noindent
    \hbox to \parindent{$^#1$\hss}\strut#2\strut\par}\hss$$
    \endgroup}

\def\L2{\ifmmode L_2\else $L_2$\fi}

\def\DeltaT{\ifmmode \Delta T\else $\Delta T$\fi}
\def\deltat{\ifmmode \Delta t\else $\Delta t$\fi}
\def\fknee{\ifmmode f_{\rm knee}\else $f_{\rm knee}$\fi}
\def\Fmax{\ifmmode F_{\rm max}\else $F_{\rm max}$\fi}
\def\solar{\ifmmode{\rm M}_{\mathord\odot}\else${\rm M}_{\mathord\odot}$\fi}
\def\Msolar{\ifmmode{\rm M}_{\mathord\odot}\else${\rm M}_{\mathord\odot}$\fi}
\def\Lsolar{\ifmmode{\rm L}_{\mathord\odot}\else${\rm L}_{\mathord\odot}$\fi}
\def\inv{\ifmmode^{-1}\else$^{-1}$\fi}
\def\mo{\ifmmode^{-1}\else$^{-1}$\fi}
\def\sup#1{\ifmmode ^{\rm #1}\else $^{\rm #1}$\fi}
\def\expo#1{\ifmmode \times 10^{#1}\else $\times 10^{#1}$\fi}
\def\,{\thinspace}
\def\lsim{\mathrel{\raise .4ex\hbox{\rlap{$<$}\lower 1.2ex\hbox{$\sim$}}}}
\def\gsim{\mathrel{\raise .4ex\hbox{\rlap{$>$}\lower 1.2ex\hbox{$\sim$}}}}

\def\simprop{\mathrel{\raise .4ex\hbox{\rlap{$\propto$}\lower 1.2ex\hbox{$\sim$}}}}
\def\deg{\ifmmode^\circ\else$^\circ$\fi}
\def\pdeg{\ifmmode $\setbox0=\hbox{$^{\circ}$}\rlap{\hskip.11\wd0 .}$^{\circ}
          \else \setbox0=\hbox{$^{\circ}$}\rlap{\hskip.11\wd0 .}$^{\circ}$\fi}
\def\arcs{\ifmmode {^{\scriptstyle\prime\prime}}
          \else $^{\scriptstyle\prime\prime}$\fi}
\def\arcm{\ifmmode {^{\scriptstyle\prime}}
          \else $^{\scriptstyle\prime}$\fi}
\newdimen\sa  \newdimen\sb
\def\parcs{\sa=.07em \sb=.03em
     \ifmmode \hbox{\rlap{.}}^{\scriptstyle\prime\kern -\sb\prime}\hbox{\kern -\sa}
     \else \rlap{.}$^{\scriptstyle\prime\kern -\sb\prime}$\kern -\sa\fi}
\def\parcm{\sa=.08em \sb=.03em
     \ifmmode \hbox{\rlap{.}\kern\sa}^{\scriptstyle\prime}\hbox{\kern-\sb}
     \else \rlap{.}\kern\sa$^{\scriptstyle\prime}$\kern-\sb\fi}
\def\ra[#1 #2 #3.#4]{#1\sup{h}#2\sup{m}#3\sup{s}\llap.#4}
\def\dec[#1 #2 #3.#4]{#1\deg#2\arcm#3\arcs\llap.#4}
\def\deco[#1 #2 #3]{#1\deg#2\arcm#3\arcs}
\def\rra[#1 #2]{#1\sup{h}#2\sup{m}}

\def\dots{\relax\ifmmode \ldots\else $\ldots$\fi}
\def\WHzsr{\ifmmode $W\,Hz\mo\,sr\mo$\else W\,Hz\mo\,sr\mo\fi}
\def\mHz{\ifmmode $\,mHz$\else \,mHz\fi}
\def\GHz{\ifmmode $\,GHz$\else \,GHz\fi}
\def\mKs{\ifmmode $\,mK\,s$^{1/2}\else \,mK\,s$^{1/2}$\fi}
\def\muKs{\ifmmode \,\mu$K\,s$^{1/2}\else \,$\mu$K\,s$^{1/2}$\fi}
\def\muKRJs{\ifmmode \,\mu$K$_{\rm RJ}$\,s$^{1/2}\else \,$\mu$K$_{\rm RJ}$\,s$^{1/2}$\fi}
\def\muKHz{\ifmmode \,\mu$K\,Hz$^{-1/2}\else \,$\mu$K\,Hz$^{-1/2}$\fi}
\def\MJysr{\ifmmode \,$MJy\,sr\mo$\else \,MJy\,sr\mo\fi}
\def\MJysrmK{\ifmmode \,$MJy\,sr\mo$\,mK$_{\rm CMB}\mo\else \,MJy\,sr\mo\,mK$_{\rm CMB}\mo$\fi}
\def\microns{\ifmmode \,\mu$m$\else \,$\mu$m\fi}

\def\muK{\ifmmode \,\mu$K$\else \,$\mu$\hbox{K}\fi}
\def\microK{\ifmmode \,\mu$K$\else \,$\mu$\hbox{K}\fi}
\def\muW{\ifmmode \,\mu$W$\else \,$\mu$\hbox{W}\fi}
\def\kms{\ifmmode $\,km\,s$^{-1}\else \,km\,s$^{-1}$\fi}
\def\kmsMpc{\ifmmode $\,\kms\,Mpc\mo$\else \,\kms\,Mpc\mo\fi}
\providecommand{\sorthelp}[1]{}